\documentclass[pra,twocolumn,showpacs,superscriptaddress]{revtex4-1}
\usepackage{epsfig,graphicx,amssymb,color,bm}
\usepackage{amsmath,float}
\usepackage[colorlinks, linkcolor=blue, urlcolor=blue, citecolor=blue, unicode=true]{hyperref}
\usepackage{setspace}
\usepackage{url}
\usepackage{subfigure} 
\usepackage{amsfonts}
\usepackage{latexsym}
\usepackage{amsbsy}
\usepackage{amsopn, amstext}
\usepackage{epstopdf}
\usepackage{threeparttable}
\UseRawInputEncoding

\newcommand{\eref}[1]{Eq.~(\ref{#1})}
\newcommand{\esref}[1]{Eqs.~(\ref{#1})}
\newcommand{\fref}[1]{Fig.~\ref{#1}}

\begin{document}

\title{Quantum Nonlinear Effect in Dissipatively Coupled Optomechanical System}
\author{Wen-Quan Yang}
\affiliation{Department of Physics, Ningbo University, Ningbo 315211, China}
\author{Wei Niu}
\affiliation{Department of Physics, Ningbo University, Ningbo 315211, China}
\author{Yong-Hong Ma}\email{myh\_dlut@126.com}
\affiliation{School of Science, Inner Mongolia University of Science and Technology, Baotou 014010, China}
\author{Wen-Zhao Zhang}\email{zhangwenzhao@nbu.edu.cn}
\affiliation{Department of Physics, Ningbo University, Ningbo 315211, China}

\begin{abstract}
A full-quantum approach is used to study the quantum nonlinear properties of a compound Michelson-Sagnac interferometer optomechanical system. 
By deriving the effective Hamiltonian, we find that the reduced system has a Kerr nonlinear term with complex coefficient, which is fully induced by the dissipative and dispersive couplings. 
And unexpectedly, the nonlinearities caused by the dissipative coupling have non-Hermitian Hamiltonian-like properties. It can preserve the quantum nature of the dispersive coupling beyond the traditional dissipation of the system. 
This protection mechanism allows the system to exhibit strong quantum nonlinear effects when $J^2 = \Delta_c \Delta_e$. 
Moreover, the additive effects of dispersive and dissipative couplings can produce strong anti-bunching effects, which exists in both strong and weak coupling conditions.
Our work may offer a new possibility for studying and producing strong quantum nonlinear effects in dissipatively coupled optomechanical system.
\end{abstract}
\maketitle
\section{Introduction}\label{sec1}
In cavity optomechanics, the interaction of optical fields with mechanical oscillators provides an excellent research platform for the study of fundamental physics and applications \cite{RevModPhys.86.1391,Kippenberg:07,doi:10.1126/science.1156032}. 
The radiation pressure or gradient forces \cite{faust2012microwave} of the conventional optomechanical interactions lead to a dispersive shift of the cavity frequency. It has been employed for squeezing of optical mode\cite{PhysRevX.3.031012,SCHNABEL20171}, entanglement between optical and mechanical mode \cite{PhysRevLett.99.250401,PhysRevResearch.4.033112}, cooling the mechanical oscillators to the quantum ground state \cite{PhysRevLett.99.073601,PhysRevLett.114.043601}, and optomechanical normal-mode splitting \cite{PhysRevA.81.053810,Weiss_2013}. 
Moreover, due to the nonlinear optomechanical interaction \cite{RevModPhys.86.1391}, many interesting applications have been demonstrated, such as, the photon blockade \cite{PhysRevLett.107.063601,PhysRevLett.107.063602,PhysRevLett.104.183601,PhysRevA.87.023809,PhysRevLett.121.153601}, Kerr nonlinearity \cite{PhysRevA.88.043826,PhysRevA.88.063854} and optomechanical induced transparency \cite{doi:10.1126/science.1195596,PhysRevLett.111.133601,PhysRevA.81.041803,teufel2011circuit}.

Generally, optomechanical interaction mainly have two different forms of coupling: the dispersive coupling \cite{RevModPhys.86.1391,PhysRevA.51.2537} and the dissipative coupling \cite{PhysRevLett.103.223901,PhysRevLett.102.207209}.
The dispersive coupling typically occurs in conventional optomechanical systems, in which the displacement of mechanical oscillator results in a shift of the resonant frequency of the optical cavity. Meanwhile, the dissipative coupling characterizes the dependence of the cavity decay rate on the displacement of the mechanical oscillator. 
The dissipative coupled linearized optomechanical system has recently attracted considerable attention, e.g., cooling the mechanical oscillators in the unresolved sideband regime \cite{PhysRevLett.102.207209,PhysRevLett.107.213604,PhysRevA.102.043520}, electromagnetically induced transparency \cite{Weiss_2013,PhysRevA.107.013524} and the squeezing of the output light \cite{PhysRevA.91.063815,Kilda_2016,PhysRevA.97.063820}.
Dissipative coupling has been expermentally achieved with different realization, such as a waveguide coupled to a microdisk resonator \cite{PhysRevLett.103.223901}, a monolayer graphene membrane in optical resonator \cite{10.1007/s00340-016-6564-z}, a tapered fiber coupled to a whispering-gallery mode \cite{10.1063/1.4922637,PhysRevLett.129.073901}, and a photonic crystal cavity \cite{PhysRevX.4.021052}. 
Among them, an attractive experiment in the optical domain with Michelson-Sagnac interferometer (MSI) has been proposed to realise dispersive and dissipative coupling, which can cool the mechanical oscillator from room temperature 293 K to 126 mK \cite{PhysRevLett.114.043601}.

In the dissipative coupled systems, the classical nonlinear effects have been well studied.
However, to our knowledge, there are very few researches discuss the property of quantum nonlinear effects, which can be characterized, in general, by the photon anti-bunching effect.
Recently, the second-order correlation of the output light \cite{Kilda_2016} and the nonreciprocal photon blockade \cite{Gao:21} were studied in dissipative coupled systems. 
We note that, the nonlinear effect discussed in Ref.~\cite{Kilda_2016} is studied under the linearization approximation, while in Ref.~\cite{Gao:21}, a single-mode approximation of the environment is made. 
These two treatments either lacks of effective description of quantum nonlinearity, or has difficulty in experimental realizations. Therefore, an experimentally realizable scheme with a general theoretical approach including the description of quantum nonlinearity is urgently needed.

In this paper, we show that the strong quantum nonlinearity can be realized in the dissipative coupling regime in the MSI optomechanical system.
The photon anti-bunching effect of the composite system, consisting of a dissipative coupled optomechanical cavity and an empty cavity, is studied by utilizing the full-quantum approach.
Our results show that the quantum nonlinear effect is highly depend on the dispersive coupling strength. 
More importantly, the interplay between dispersive and dissipative couplings may produce strong anti-bunching (blockade) even in the weakly coupling regime.
We also find the optimal parameter condition for producing strong blockade effect, i.e., $\Delta_c \Delta_e=J^2$.
Moreover, the dissipative coupling plays a key role in protecting the quantum nature of the system.
When the gain of the dissipative coupling strength equals to the loss brought by the conventional dissipation, the dissipative coupling will balance the effect of system loss. Then the system can be approximately described as a closed system, and the quantum nonlinear effect is highlighted. 

The paper is organized as follows. 
In Sec. II we introduce the system Hamiltonian and derive the effective Hamiltonian. 
In Sec. III, both numerical and analytical methods are used to study the second-order correlation function. 
Finally, we conclude and discuss the feasibility of experiment in Sec. IV.

\section{Model and HAMILTONIAN}\label{sec2}
\begin{figure}
\centering
\includegraphics[width=0.5\textwidth]{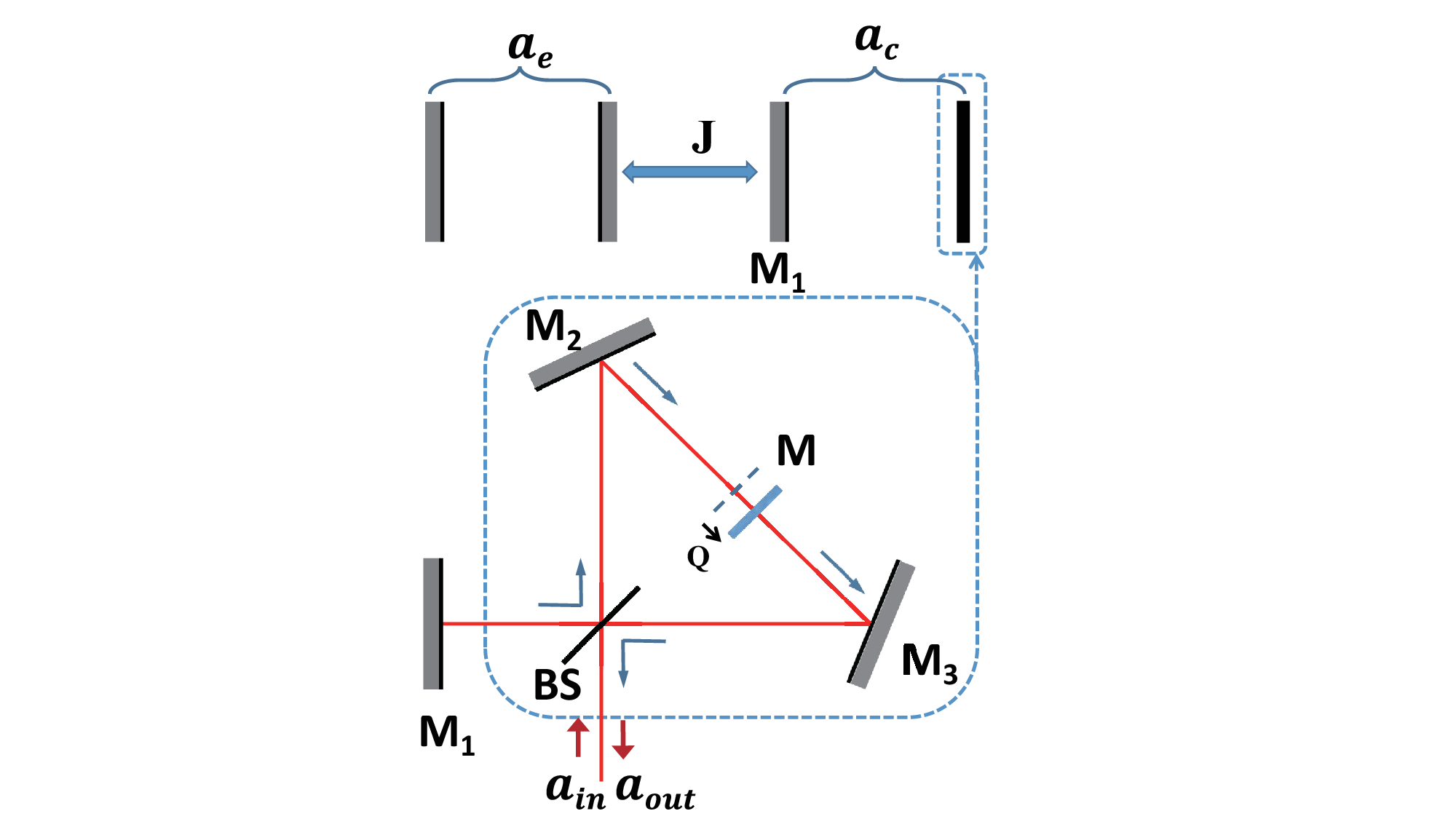}
    \caption{(color online). The sketch of Michelson-Sagnac interferometer \cite{PhysRevLett.107.213604} with a movable  membrane between the
mirror $M_2$ and $M_3$ as the effective MSI mirror (the area surrounded by blue dotted line) combined with $M_1$  into a  Fabry-P$\mathrm{\acute{e}}$rot cavity which coupled to an empty cavity with the tunneling coupling strength $J$. 
}
    \label{fig:1}
\end{figure}
A compound MSI optomechanical model \cite{PhysRevA.81.033849} that enables both the dispersive and dissipative coupling is shown in Fig.~\ref{fig:1}. 
Our model contains a standard optical cavity and a MSI optomechanical system which can realize the dispersive and dissipative coupling.
In MSI optomechanical system, $\text{M}_1$, $\text{M}_2$ and $\text{M}_3$ are perfect total reflection mirrors, while the movable membrane $\text{M}$ is a semitransparent membrane. 
As is shown in the diagram, the beam splitter (BS) in the optical path is used to control the input and output of the MSI-cavity.
The reflectivity $R$ and transmissivity $T$ of the BS can determine the magnitude of the dispersive and dissipative coupling \cite{PhysRevA.88.023809}, and in extreme situations we can even obtain a purely dissipative coupling \cite{PhysRevLett.107.213604}. 
The effective MSI-cavity length can be obtained by halving the length of the Sagnac mode $M_1\to BS\to M_2\to M_3 \to BS$ \cite{PhysRevLett.107.213604}.
The displacement of the membrane $M$ can change the cavity length as well as the the interference optical path of the MSI. 
Thus the membrane will influence the cavity frequency and the input-output relationship, and hence arise dispersive and dissipative coupling.
The total Hamiltonian of the model is read \cite{PhysRevLett.102.207209} ($\hbar=1$)
\begin{eqnarray}\label{eq.1}
\hat{H}_T&=&\omega_{c}\hat{a}_c^{\dag}\hat{a}_c+\omega_{e}\hat{a}_{e}^{\dag}\hat{a}_e+\frac{1}{2}\omega_{m}(\hat{Q}^2+\hat{P}^2)\\&&+\left[A\kappa_c\hat{a}_c^{\dag}\hat{a}_c+\frac{B}{2}\hat{H}_{int}\right]\hat{Q}+\hat{H}_{int}+\hat{H}_{diss}\nonumber\\&&+J(\hat{a}_{c}^{\dag}\hat{a}_e+\hat{a}_c\hat{a}_{e}^{\dag})+\sum_{j=c,e}\epsilon_{j}(\hat{a}_{j}^{\dag}e^{-i\omega_{d}t}+h.c.).\nonumber
\end{eqnarray}
The first three terms in \eref{eq.1} represent the free energy of the optomechanical cavity, empty cavity and the mechanical oscillator, respectively.
$\hat{Q}=\hat{x}/x_{zpf}$ and $\hat{P}=\hat{p}/(m\omega_m x_{zpf})$ are the dimensionless displacement and momentum operators, respectively.
$\hat{x}=x_{zpf}(\hat{b}+\hat{b}^{\dag})$ is the displacement of mechanical oscillator, and $x_{zpf}=(2m\omega_m)^{-\frac{1}{2}}$ denotes the size of the zero-point fluctuations.
The fourth term represents the total cavity-mechanical coupling.
The parameters $A$ and $B$, which can be derived from the frequency and dissipative shifts, denote the weights of the dispersive and dissipative couplings, respectively.
Where the corresponding optical frequency shift per zero-point fluctuation is $g_\omega=\frac{\partial\omega_c(Q)}{\partial x}x_{zpf}=A\kappa_c$ (dispersive coupling), and dissipation shift per zero-point fluctuation is $g_\kappa=\frac{\partial\kappa_c(Q)}{\partial x}x_{zpf}=B\kappa_c$ (dissipative coupling). $\omega_c(Q)$ and $\kappa_c(Q)$ denote displacement modified frequency and dissipation, respectively.
The fifth term represents the cavity-bath interaction with $\hat{H}_{int}=i\sqrt{\frac{\kappa_c}{2\pi\rho}}\sum_{q}(\hat{a}_c^{\dag}\hat{b}_{q}-\hat{b}_{q}^{\dag}\hat{a}_c)$  \cite{zoller1997quantum,RevModPhys.82.1155}.
$\hat{H}_{diss}$ represents the commonly bath of the empty-cavity and oscillator as well as the system-bath interactions \cite{PhysRevA.31.3761}. 
The second last term represents the tunneling coupling of optomechanical cavity and empty cavity. 
The last term denotes the laser driving, which can be introduced by the BS as a semi-classical input to the MSI in our model.
\par With displacement-modified input-output relation in Markovian regime \cite{PhysRevLett.107.213604,PhysRevA.31.3761}, the Hamiltonian can be reduced by equating the environmental operators $\sqrt{\frac{\kappa_c}{2\pi\rho}}\sum_{q}\hat{b}_q$ in the system-environment interaction to $\sqrt{\kappa_c}\hat{a}_{c,in}-\frac{\kappa_c}{2}\hat{a}_{c}\left(1+\frac{g_\kappa}{2\kappa_c}\hat{Q}\right)$ \cite{Weiss_2013}. 
After eliminating the bath of MSI-cavity and rotating frame operation, the reduced Hamiltonian  is then given by \cite{PhysRevA.107.013524,PhysRevLett.107.213604}

\begin{eqnarray}\label{eq.2}
\hat{H}_R&=&-\Delta_c(Q)\hat{a}_c^{\dag}\hat{a}_c-\Delta_{e}\hat{a}_{e}^{\dag}\hat{a}_e+\frac{1}{2}\omega_{m}(\hat{Q}^2+\hat{P}^2)\\
&&+J(\hat{a}_{c}^{\dag}\hat{a}_e+\hat{a}_c\hat{a}_{e}^{\dag})+H_{diss}+\sum_{j=c,e}\epsilon_{j}(\hat{a}_{j}^{\dag}+\hat{a}_{j}),\nonumber
\end{eqnarray}
where $\Delta_c(Q)=\omega_d-\omega_c(Q)$ is the displacement-modified detuning.
$\Delta_c=\omega_d-\omega_c$ and  $\Delta_e=\omega_d-\omega_e$ are the driving-detuning of the optomechanical cavity and empty cavity, respectively.
Under the first-order approximation (see the Appendix~\ref{app.1}), we derive the expressions $\omega_c(Q)=\omega_c+g_{\omega}\hat{Q}$ and $\kappa_c(Q)=\kappa_c+g_{\kappa}\hat{Q}$
with  $\kappa_c(Q)$ the displacement-modified dissipation.
Note that the modified input-output relation $\hat{a}_{c,out}-\hat{a}_{c,in}= \sqrt{\kappa_c(Q)} \hat{a}_c$ \cite{PhysRevLett.107.213604,PhysRevA.31.3761} is well-adapted method to investigate the quantum dynamic of the dissipative coupled system. 
According to \eref{eq.2}, the nonlinear quantum Langevin equations  in Markovian regime are given by \cite{PhysRevA.105.013503,Mehmood_2020}
\begin{subequations}
\begin{eqnarray}
\dot{\hat{Q}}&=&\omega_m\hat{P},\label{eq.3a}\\
\dot{\hat{P}}&=&-g_\omega\hat{a}_c^{\dag}\hat{a}_c-\omega_m\hat{Q}-\frac{ig_\kappa}{2\sqrt{\kappa_c}}(\hat{a}_c^{\dag}\hat{a}_{c,in}-\hat{a}_{c,in}^{\dag}\hat{a}_c)\nonumber\\&&-\gamma\hat{P}+\hat{\xi},\label{eq.3b}\\ 
\dot{\hat{a}}_{c}&=&\left[i(\Delta_{c}-g_\omega\hat{Q})-\frac{\kappa_{c}}{2}(1+\frac{g_\kappa}{\kappa_c}\hat{Q})\right]\hat{a}_{c}-iJ\hat{a}_e\nonumber\\&&+\sqrt{\kappa_c}\left(1+\frac{g_\kappa}{2\kappa_c}\hat{Q}\right)\hat{a}_{c,in}+\epsilon_c,\label{eq.3c}\\
\dot{\hat{a}}_{e}&=&(i\Delta_{e}-\frac{\kappa_{e}}{2})\hat{a}_{e}-iJ\hat{a}_c+\sqrt{\kappa_e}\hat{a}_{e,in}+\epsilon_e.
\end{eqnarray}
\end{subequations}
In our discussion, we mainly focus on optical modes.
Since the experiment reported decay rate of the mechanical oscillator is much smaller than that of the cavity \cite{RevModPhys.86.1391}, we can safely use the quantum steady state solution of mechanical mode to solve the optical modes under conditions of long time evolution, in which
\begin{eqnarray}
\hat{Q}&=&\frac{-g_\omega\hat{a}_c^{\dag}\hat{a}_c-\frac{ig_\kappa}{2\sqrt{\kappa_c}}(\hat{a}_c^{\dag}\hat{a}_{c,in}-\hat{a}_{c,in}^{\dag}\hat{a}_c)+\hat{\xi}}{\omega_m}.
\label{eq.4}
\end{eqnarray}
Bringing \eref{eq.4} back to \eref{eq.3c}, the mechanical mode can be reduced and thus obtained the dynamical equations for the long-time evolution of the optical modes,
\begin{subequations}
\begin{eqnarray}
\dot{\hat{a}}_{c}&=&(i\Delta_{c}-\frac{\kappa_{c}}{2})\hat{a}_c+\frac{2ig_{\omega}^2+g_\kappa g_\omega}{2\omega_m}\hat{a}_c^{\dag}\hat{a}_c\hat{a}_{c}\nonumber\\&&-iJ\hat{a}_L+\sqrt{\kappa_c}\hat{\xi}_{eff}+\epsilon_c,\\
\dot{\hat{a}}_{e}&=&(i\Delta_{e}-\frac{\kappa_{e}}{2})\hat{a}_{e}-iJ\hat{a}_c+\sqrt{\kappa_e}\hat{a}_{e,in}+\epsilon_e.
\end{eqnarray}\label{eq.5}
\end{subequations}
where $\hat{\xi}_{eff}$ denotes the effective noise operator: 
\begin{eqnarray}
\hat{\xi}_{eff}&=& \frac{ig_\kappa^2-2g_\omega g_\kappa}{4\omega_m\kappa_c}(\hat{a}_c^{\dag}\hat{a}_c\hat{a}_{c,in}-\hat{a}_{c,in}^{\dag}\hat{a}_c\hat{a}_c)+\hat{a}_{c,in}\nonumber\\&&-\frac{ig_\kappa^2}{4\omega_m \kappa_c^{3/2}}(\hat{a}^{\dag}_c\hat{a}_{c,in}\hat{a}_{c,in}-\hat{a}_{c,in}^{\dag}\hat{a}_{c,in}\hat{a}_c)\nonumber\\&&+\frac{g_\kappa}{2\omega_m \kappa_c}\hat{\xi}\hat{a}_{c,in}-\frac{(2ig_\omega+g_\kappa)}{2\omega_m\sqrt{\kappa_c}}\hat{\xi}\hat{a}_c\nonumber\\&&-\frac{g_\kappa g_\omega}{2\omega_m \kappa_c}\hat{a}_{c}^{\dag}\hat{a}_c\hat{a}_{c,in}.
\label{eq.6}
\end{eqnarray}
By inverting the dynamical \esref{eq.5}, the effective Hamiltonian is obtained
\begin{eqnarray}
 \hat{H}_{eff}&=&\sum_{j=c,e}\left[-\Delta_{j}\hat{a}_{j}^{\dag}\hat{a}_j+\epsilon_{j}(\hat{a}_{j}^{\dag}+\hat{a}_{j})\right]+J(\hat{a}_{c}^{\dag}\hat{a}_e+\hat{a}_c\hat{a}_{e}^{\dag})\nonumber\\&&-\frac{(2g_{\omega}^2-ig_\kappa g_\omega)}{2\omega_m}\hat{a}_c^{\dag}\hat{a}_c^{\dag}\hat{a}_c\hat{a}_{c}.
 \label{eq.7}
\end{eqnarray}
This effective Hamiltonian describes the square-coupled nonlinear form of the optical modes in the dissipative coupled system.  According to the last term in \eref{eq.7}, the dispersive coupling strength $g_\omega\neq 0$ is a prerequisite for the nonlinear effect of the system. Moreover,  this dispersive coupling affects both the amplitude and phase of the light (corresponding to the imaginary and real parts of the nonlinear term), whereas dissipative coupling affects only the amplitude.
It is worth noting that this dissipative nonlinear effect is the opposite of the loss effect of the cavity field to the environment. 
Again, it is visible from the comparison of the sign of $\frac{\kappa_c}{2} \hat{a}_c$ and $\frac{g_\kappa g_\omega \hat{N}_c}{2\omega_m}\hat{a}_{c}$ in \esref{eq.5}. 
This provides us with the possibility that if we choose the suitable strength of the dissipative nonlinearity is able to resist the loss from the environment and thus protect the quantum property of the system.
This effect of resisting noise is very similar to the processing of non-Hermitian Hamiltonian \cite{PhysRevA.103.053508} and is verified in \fref{fig:6} of our discussion later.

Analyzing the property of the effective noise $\hat{\xi}_{eff}$  is necessary to characterize the effective Hamiltonian and thus investigate the nonlinear effects of the optical modes in the system more reasonably.
In Markov regime, the interaction between the environment and the system is quite weak. 
Moreover, under the fact that the environment with much larger degrees than the system. 
Thus, the quantum fluctuations of the environment itself can be neglected when considering the average effect of the environment on the system. 
In the solution of the Hamiltonian, we can utilize a semi-classical treatment of the effect of the environment as an effective temperature. Especially in the long-time case (steady state), as the system and the environment reach thermal equilibrium, this treatment provides a  reasonable description of the behaviour of the thermal environment on the system. 
All the approximations used in our efficientisation as well as in the later discussion are carried within this premise.
The corresponding effective noise is read,
\begin{eqnarray}
\hat{\xi}_{eff}& \rightarrow &\sqrt{\langle\hat{\xi}_{eff}^{\dag}\hat{\xi}_{eff}\rangle}\nonumber\\&=&\left[\frac{g_\kappa^4+2ig_\omega g_\kappa
 ^3}{16\omega_m^2\kappa_c^{3/2}}\Big(\sqrt{6(\bar{N}^3-3\bar{N}^2+2\bar{N})}\nonumber\right.\\&&\left.-\bar{N}\sqrt{\Bar{N}-1}\Big)+\frac{4g_\omega^2g_\kappa^2+g_\kappa^4}{16\omega_m^2\kappa_c}(\bar{N}^2-\bar{N})\nonumber\right.\\&&\left.+\frac{(g_\kappa^2+4g_\omega^2)}{4\omega_m^2}\bar{N} \bar{n}_{th}\right]^\frac{1}{2}/\sqrt{\kappa}_c,
 \label{eq.8}
\end{eqnarray}
where $\bar{N}=\langle \hat{a}_c^{\dag}\hat{a}_c\rangle$ means average photon number.  
$\bar{n}_{th}=(e^{\omega_m/k_B T_b}-1)^{-1}$ represents the thermal phonon number of the mechanical oscillator with the Boltzmann constant $k_B$ and the environmental temperature $T_b$ of the mechanical oscillator.

\begin{figure}[htp!]
    \centering
\includegraphics[width=9cm]{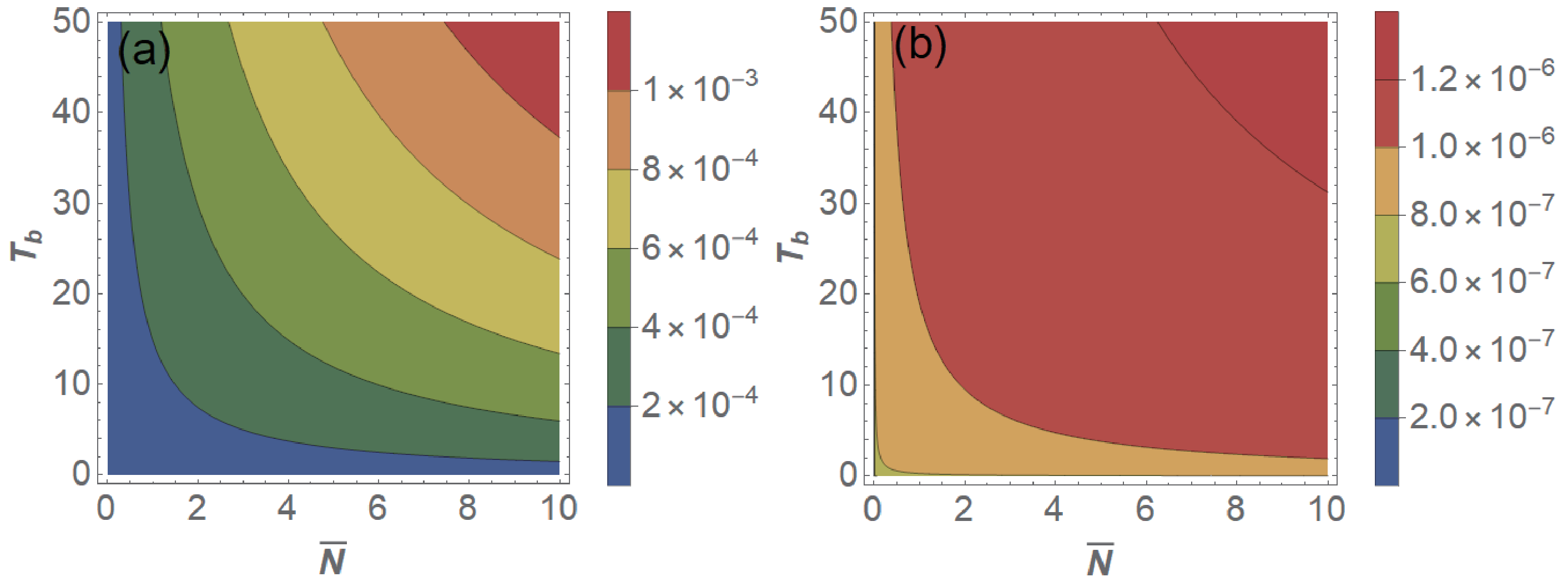}
    \caption{(Color online) (a) Effective noise as a function of the photon number $\bar{N}$ and phonon temperature $T_b(K)$. (b) Photon temperature $T_a(K)$ as a function of the photon number $\bar{N}$ and phonon temperature $T_b$. The other parameters are $\omega_m=10^6 Hz, \kappa=5\times10^3 Hz, g_\omega=200Hz, g_\kappa=500Hz.$ }
    \label{fig:2}
\end{figure}
The effective noise and the photon temperature of the system with the different photon number $\bar{N}$ and phonon temperature $T_b(K)$ are illustrated in \fref{fig:2}.
Obviously, the lower the average photon number or the phonon temperature is helpful to obtain a less environmental excitation number (As shown in
Fig.~\ref{fig:2}(a)) and photon temperature  (As shown in
Fig.~\ref{fig:2}(b)).  
Note that the effective environmental excitation number is almost 4 orders of magnitude smaller than the average photon number.  The effective environmental temperature is 7 to 8 orders of magnitude smaller than the real temperature  (e.g. the temperature is $10$K and the average photon number is $2$ corresponding to the effective environmental temperature is $8\times 10^{-7}$K). As a result, our system has much stronger temperature robustness. Compared with the general experimental phonon temperatures \cite{RevModPhys.86.1391}, the effective environmental temperature is close to zero as long as the driving field is weak enough to ensure a low level of average intracavity photon number.  

\section{Second order correlation function for optical mode}\label{sec3}
The second-order correlation function is a very effective formula to evaluate quantum nonlinearity, and is easily detected in experiment \cite{nphys.7.879}. To investigate the properties of the optical quantum nonlinear effects in dissipative coupled systems, both numerical and analytical methods are used to solve the second-order correlation function for specific analysis. For simplicity, the state truncation method \cite{PhysRevA.83.021802} is used to obtain the analytical solution.
\par
\begin{figure}
\centering
    \includegraphics[trim=200 10 350 10,clip,width=8.5cm]{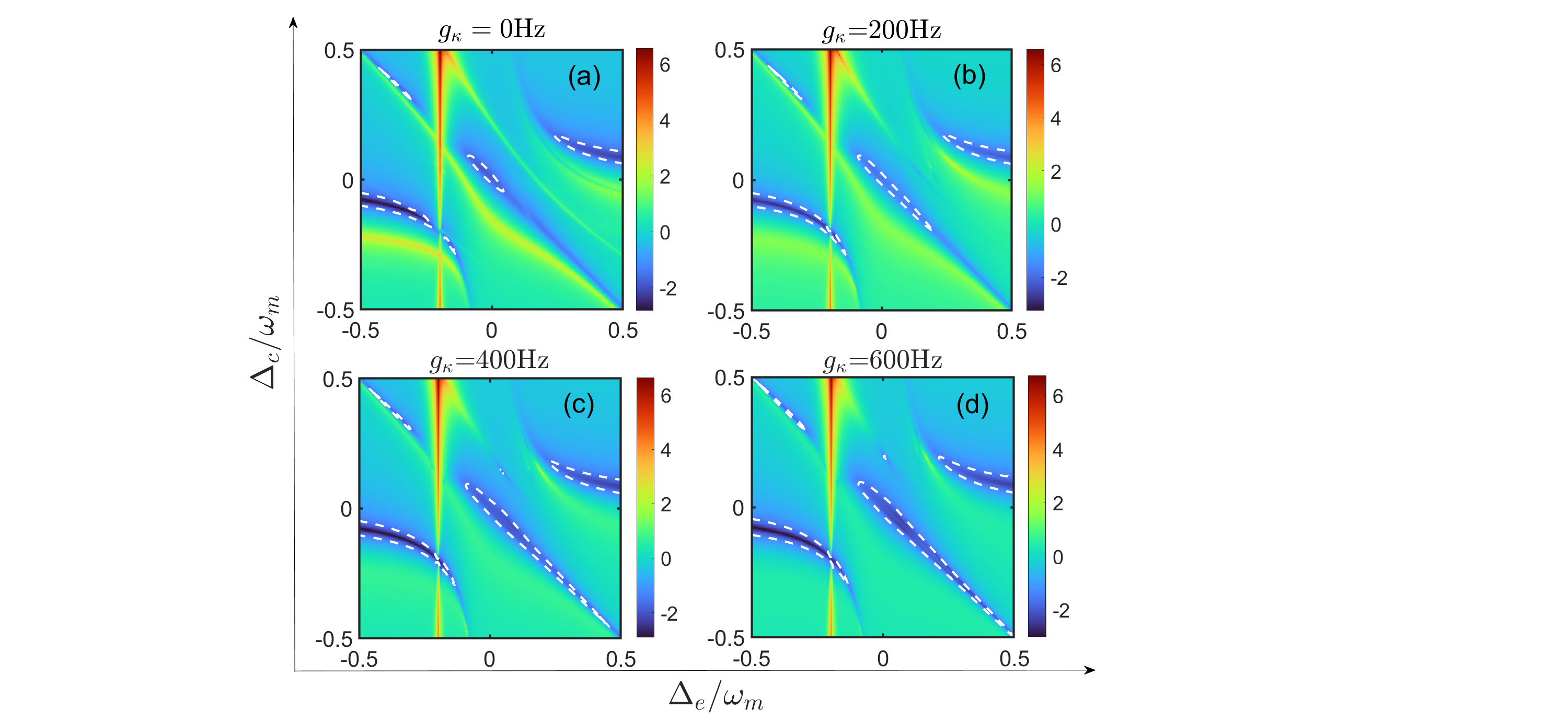}
    \caption{(Color online) Zero-time second-order correlation $g^2(0) $ as a function of the detuning $\Delta_c/\omega_m$ and $\Delta_e/\omega_m$ in optomechanical cavity. Where $g_\omega=400Hz$ does not change and $g_\kappa$ is 0, 200, 400 and 600 Hz respectively. The white dotted line represents the value of $log_{10}[g^2(0)] = -1.5$. The other parameters are $\omega_m=10^6 Hz, \kappa=5\times10^3 Hz, J=2\times10^5 Hz, \epsilon_c=\epsilon_e=5\times10^3 Hz.$ }
    \label{fig:4}
\end{figure}

\begin{figure}[htp!]
    \centering
    \includegraphics[trim=200 10 350 10,clip,width=8.5cm]{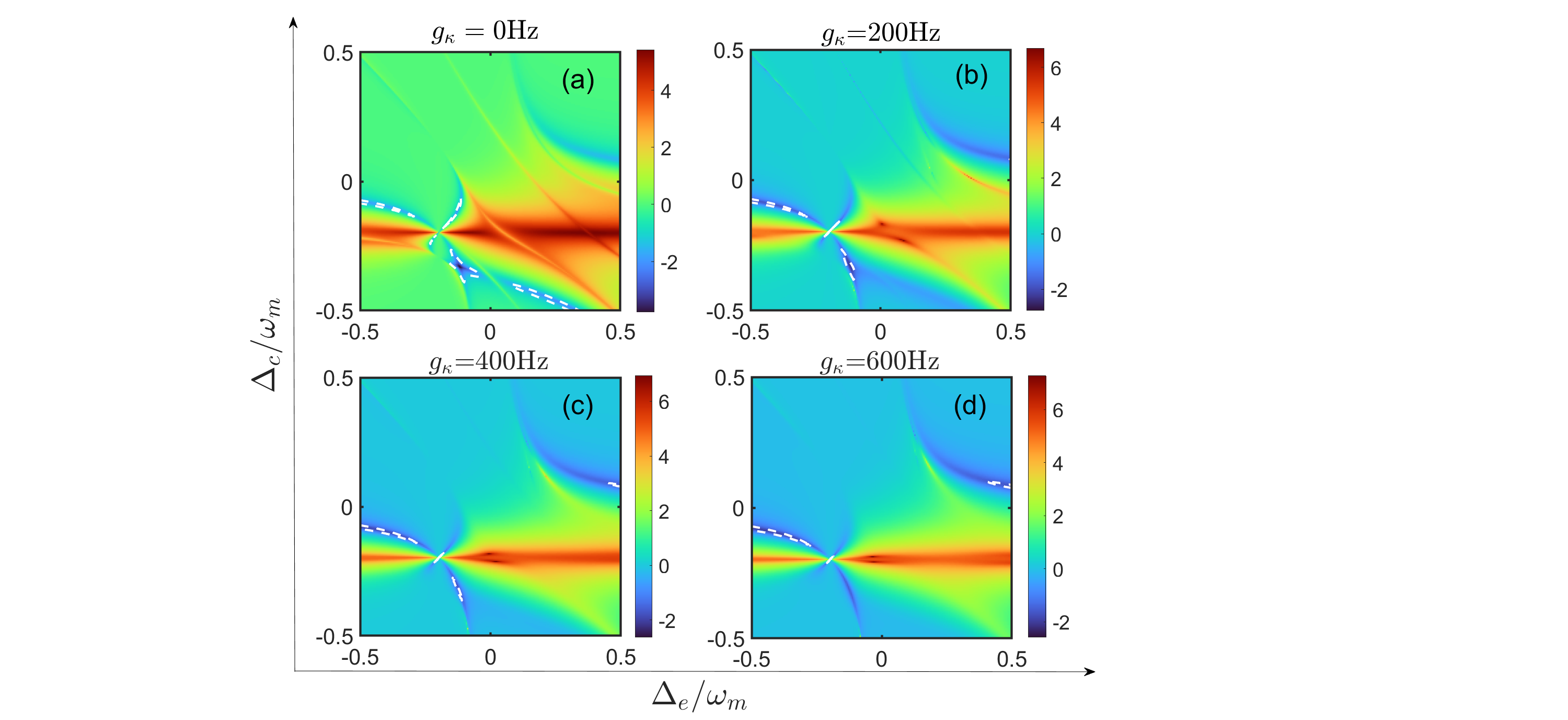}
    \caption{(Color online) Zero-time second-order correlation $g^2(0) $ as a function of the detuning $\Delta_c/\omega_m$ and $\Delta_e/\omega_m$ in empty cavity. Where $g_\omega=400Hz$ does not change and $g_\kappa$ is 0, 200, 400 and 600 Hz, respectively. The white dotted line represents the value of $log_{10}[g^2(0)] = -1.5$. The other parameters are $\omega_m=10^6 Hz, \kappa=5\times10^3 Hz, J=2\times10^5 Hz, \epsilon_c=\epsilon_e=5\times10^3 Hz.$ }
    \label{fig:5}
\end{figure}

 In the weak pumping conditions $C_0\gg C_{j,ce}\gg C_{jj}, \{j=c,e\}$ and after utilising the truncation method, the general optical quantum state  is expressed as,
\begin{eqnarray}
|\psi(t)\rangle&=&C_0(t)|00\rangle+C_{c}(t)|10\rangle+C_{e}(t)|01\rangle\nonumber\\&&+C_{ce}(t)|11\rangle+C_{cc}(t)|20\rangle+C_{ee}(t)|02\rangle,
\end{eqnarray}
where $C_{m,n}$ and $|mn\rangle$ represent the probability amplitude and the photon number state (corresponding to OM cavity (c) and empty cavity (e)), respectively. 
Besides, we use second-order correlation function with zero-time delay to descript the statistical properties of photons
\begin{eqnarray}
 g_j^{(2)}(0)&=&\frac{\langle \hat{a}_j^{\dag}\hat{a}_j^{\dag}\hat{a}_j\hat{a}_j\rangle}{\langle \hat{a}_j^{\dag}\hat{a}_j\rangle^2} , \quad \quad  j=c,e
\end{eqnarray}
$g^{(2)}(0)<1$ means the appearance of the quantum nonlinear effect, and the intracavity photons show an anti-bunching effect. Especially,  this effect is strongest and photon blockade is achieved when $g^{(2)}(0)=0$.
Instead, when $g^{(2)}(0)>1$, the intracavity photons show a bunching effect and the system exhibits classical behaviour.
By setting $\kappa_c=\kappa_e=0$, the analytic solution of $g^{(2)}(0)$ can be obtained as (details see Appendix \ref{app.2})
\begin{subequations}
\begin{eqnarray}
g^{(2)}_c(0)&=&\left| \frac{2 \omega_mKD_J}{ f_A}\right|^2,\label{eq.11a}\\
g^{(2)}_e(0)&=& \left|\frac{D_J \left[Gf_B+2\omega_mK(J+\Delta_c)^2\right] }{ (J+\Delta_c)^2f_A}\right|^2, \label{eq.11b} 
\end{eqnarray}\label{eq.11}
\end{subequations}
where $f_A=(2\omega_m K+G)D_J-G\Delta_e^2$, $f_B=2J^2+2JK+\Delta_cK$,  $D_J=J^2-\Delta_c\Delta_e$, $K=\Delta_c+\Delta_e$ and $G=2g_\omega^2-i g_\kappa g_\omega$.
According to \esref{eq.11a},  when $D_J=0$ corresponding to $g_j^{(2)}(0) = 0$, the coupling strength and detuning between the optical cavities satisfy $J^2=\Delta_c\Delta_e$, thus the photon blockade effect appears. 
In this case, a fixed $J$ will result in a hyperbolic relation between the  $g_j^{(2)}(0)$ and $\Delta_j$. In addition, $K = 0$ also causes $g_c^{(2)}(0)=0$ (corresponding to $\Delta_c=-\Delta_e$),  while can not causes $g_e^{(2)}(0)=0$. When $J = -\Delta_c$ corresponding to $g_e^{(2)}(0) \rightarrow \infty$, the empty cavity exhibits a strong bunching effect. These properties are also shown in \fref{fig:4} and \ref{fig:5} in the following discussion.

To correctly account for the nonlinear effects character of the system, we introduce the quantum master equation \cite{PhysRevA.91.063836} for the system density matrix,
\begin{eqnarray}
\dot{\rho}&=&-i[\hat{H},\rho]+\sum_{j=c,e}\frac{\kappa_j}{2}\mathcal{D}[\hat{a}_j]\rho + \frac{\gamma}{2}(n_{th}+1)\mathcal{D}[\hat{b}]\rho\nonumber\\&&+\frac{\gamma}{2}n_{th}\mathcal{D}[\hat{b}^{\dag}]\rho. \label{meq}
\end{eqnarray}
$\kappa_j$ and $\gamma$ are the cavity and mechanical energy decay rates, respectively. 
$ \mathcal{D}[\hat{o}]\rho=2\hat{o}\rho\hat{o}^{\dag}-\hat{o}^{\dag}\hat{o}\rho-\rho\hat{o}^{\dag}\hat{o}$ is the Lindblad dissipation superoperator that accounting for losses to the environment.
Then, we can derive the corresponding numerical solutions of \esref{meq} to analyse the effect of the parameters on the second-order correlation function.
\begin{figure}[htp!]
    \centering
    \includegraphics[trim=100 100 200 50,clip,width=8.5cm]{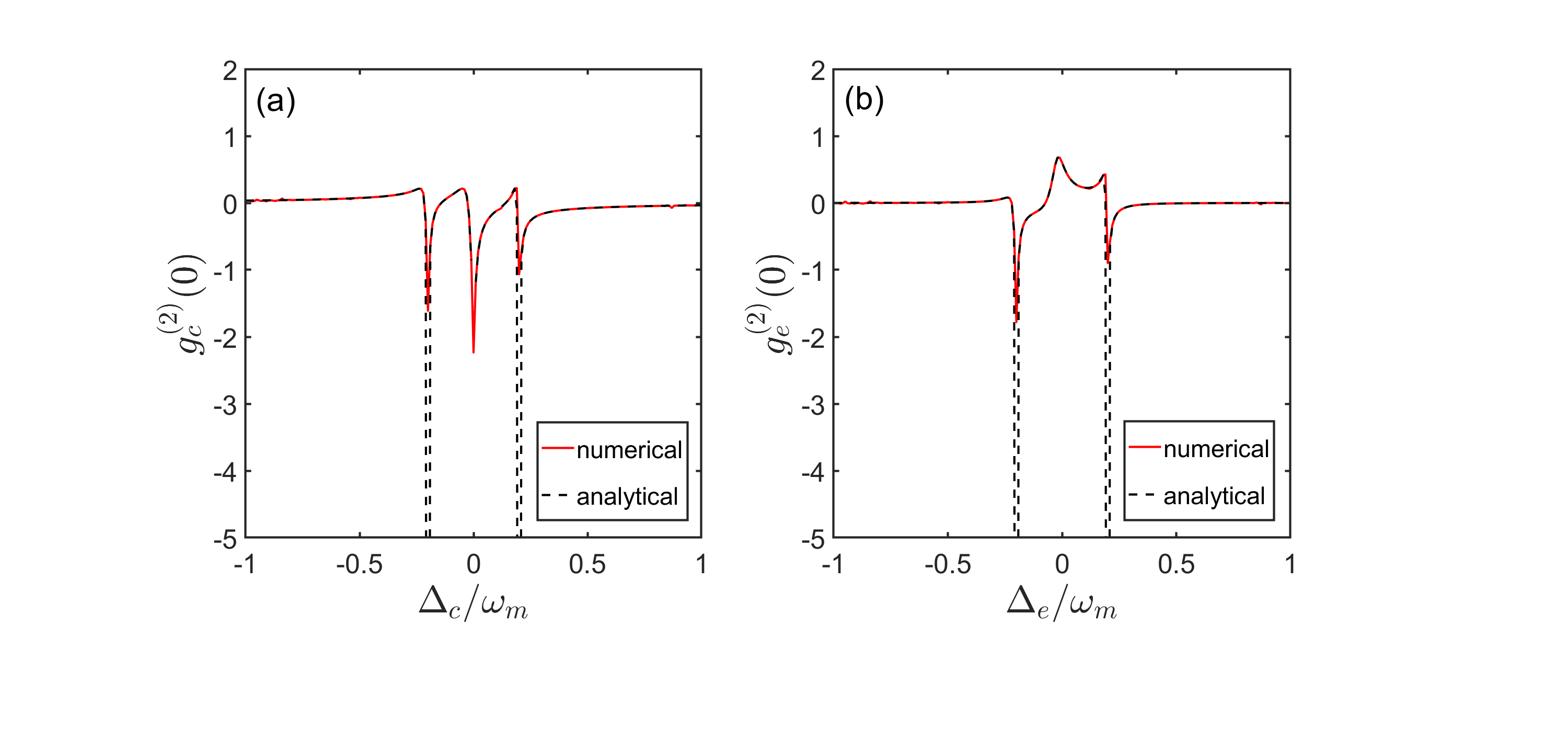}
    \caption{(Color online) The comparison of the numerical and analytical solutions in the optomechanical cavity (a) and auxiliary cavity (b) with $\Delta_c=\Delta_e=\Delta$. The vertical coordinate is zero-time second-order correlation $log_{10}g^{(2)}(0)$. The other parameters are $\omega_m=10^6 Hz, \kappa=5\times10^3 Hz, g_\omega=200Hz,g_\kappa=500Hz,J=2\times10^5 Hz, \epsilon_c=\epsilon_e=5\times10^3 Hz.$ }
    \label{fig:3}
\end{figure}
The comparison of the analytical   and numerical solutions are shown in \fref{fig:3}.
These two results are well matched in the figures.
Surprisingly, there is a sharp decrease in both the analytic and numerical solutions at $\Delta/\omega_m = -J/\omega_m = -0.2$, which corresponds to $D_J = 0$ (as we have analysed in our analytical solution), i.e., $g^{(2)}(0) = 0$.
At this singularity, assumption of $\kappa = 0$ in analytical solution causes the stronger anti-bunching effect from the analytic solution than from the numerical ones, corresponding to the black-dashed line is much deeper than the red-solid line. Moreover, $g^{(2)}_c(0)$ from both the numerical and analytical solutions can reach their minimums in \fref{fig:3}(a), which can be explained by the previous analytical solution i.e., $g^{(2)}_c(0)=0$ with $K=0$. 
When $\Delta_j/\omega_m>0.2$, we can see that $g^{(2)}_c(0)$ both from numerical and analytical solutions converge to 1, which means that the light is in coherent state for the large detuning case.

The zero-time second-order correlation of photons in the optomechanical cavity and the empty cavity as a function of the two cavities' detuning are shown in \fref{fig:4} and \ref{fig:5}, respectively. 
As shown in \fref{fig:4}, the near-photon blockade region (the region surrounded by the white dashed line) matches well with the analytical solution, corresponding to the hyperbolic region and the linear region.
The simplified expressions between $\Delta_c$ and $\Delta_e$ can be obtained according to \esref{eq.11}: $\Delta_c=J^2/\Delta_e$ and  $\Delta_c=-\Delta_e$  for the hyperbolic region and for the linear region, respectively.
Except for above special conditions, rapid enhancement of $g_c^{(2)}(0)$ will cause the appearance of classical effects for the cavity .
Especially when $-\Delta_e = J$, \fref{fig:4} shows a strong bunching effect.
This special condition can be understood from the effective Hamiltonian in \eref{eq.7}, when $-\Delta_e=J$, the detuning energy of the empty-cavity is exactly the same as the energy of the BS interaction.
Therefore, the energy of the empty-cavity can be converted into the OM-cavity through a resonance-like effect.
Due to the condition $J\gg \{g_\kappa,g_\omega \}$, the converted energy can eliminate the blockade effect due to nonlinear energy shift ($\propto n^2$) and excite the photons in the OM-cavity, thus to exhibit a bunching effect.

Comparing four plots in \fref{fig:4} or \ref{fig:5}, it shows that $g_\kappa$ has an optimising effect on $g_c^{(2)}(0)$. From \fref{fig:4}, although the increase of $g_\kappa$ can not result a significant decrease of $g_c^{(2)}(0)$, while can increase the range of parameters in the near-photon blockade region. 
It can be seen from the expression $g^{(2)}_c(0) \propto \frac{1}{|(D_J-\Delta_e^2)G+2\omega_mKD_J|^2}$. Obviously, $g^{(2)}_c(0)$ is decreasing with the increase of $|G|$ ($|G|$ increases with $g_\kappa$).
Note that this increasing trend is relatively saturated when $g_\kappa$ exceeds 400Hz.
A similar conclusions can be obtained in the empty-cavity, as shown in \fref{fig:5}, where the region of the minimum value of the second-order correlation function (the blue region in the figure) exhibits a clear hyperbolic function characteristic.
Differently, the near-photon blockade region is gradually becoming smaller as $g_\kappa$ increases, which is  widely divergent to Fig. \ref{fig:4}, due to the different expression $g^{(2)}_e(0)\propto \left|\frac{D_J f_B G+X}{ (J+\Delta_c)^2(D_J-\Delta_e^2)G+Y}\right|^2$ ($X$ and $Y$ are constant), and $|D_J f_B|> |(J+\Delta_c)^2(D_J-\Delta_e^2)|$, thus  $g^{(2)}_c(0)$ is increasing with the increase of $|G|$. 
Therefore, $g_\kappa$ has a negative damping of the photon blockade effect in the empty cavity. 
\begin{figure}[htp!]
    \centering
\includegraphics[trim=0 60 0 0,clip,width=9cm]{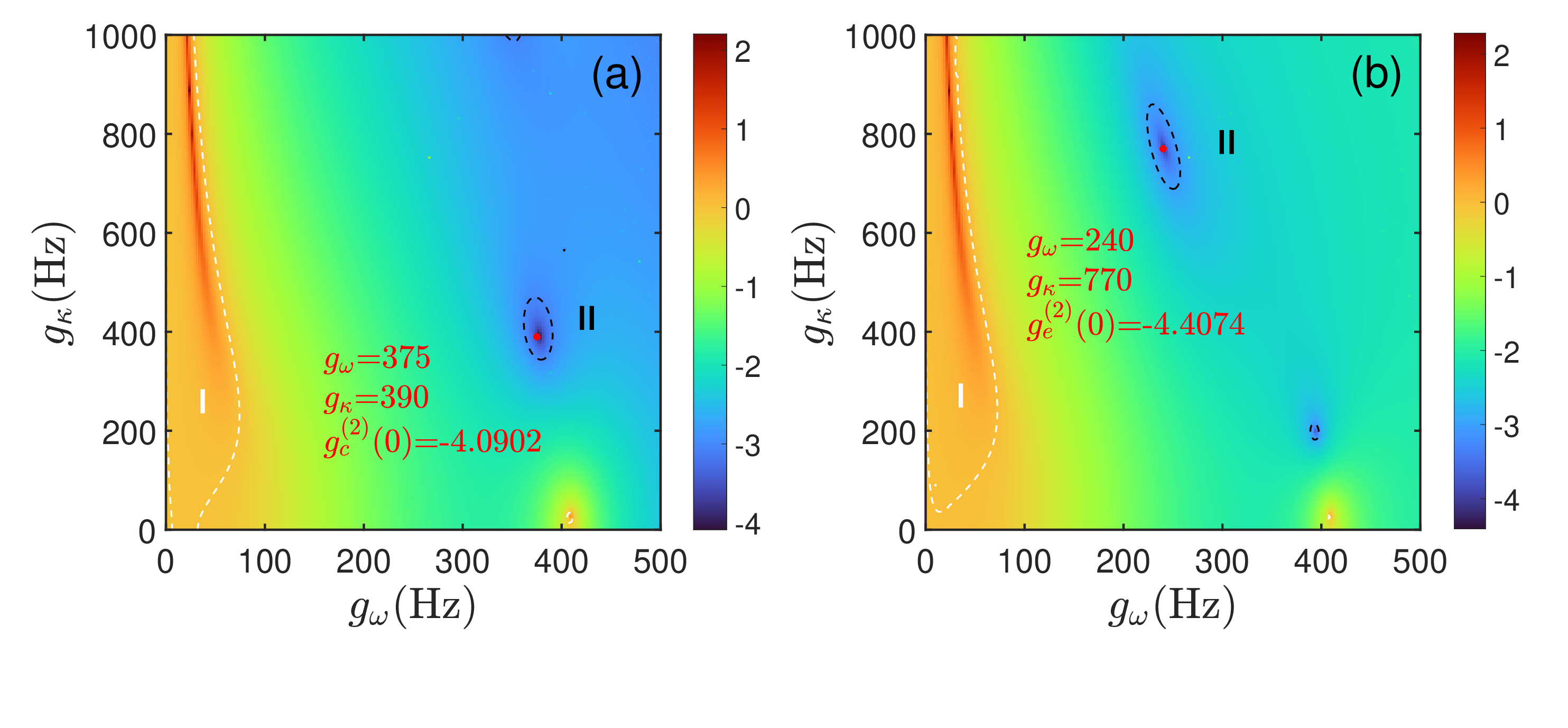}
    \caption{(Color online) Zero-time second-order correlation $g^{(2)}(0) $ as a function of the dispersive coupling strength $g_\omega$ and dissipative coupling strength $g_\kappa$ in optomechanical cavity (a) and empty cavity (b).The blank dotted line represents the value of $\log_{10}[g^{(2)}(0)]= -3$, the white dotted line represents the value of $\log_{10}[g^{(2)}(0)] = 0$, and the red point represent the minimum of the $g^{(2)}(0)$(It's size and coordinatesis are also shown in the figure). The other parameters are $\omega_m=10^6 Hz, \kappa=5\times10^3 Hz, \Delta_c=\Delta_e=-2\times10^5 Hz, J=2\times10^5 Hz, \epsilon_c=\epsilon_e=5\times10^3 Hz.$ }
    \label{fig:6}
\end{figure}
\begin{figure}[htp!]
    \centering
    \includegraphics[trim=10 10 0 20,clip,width=0.5\textwidth]{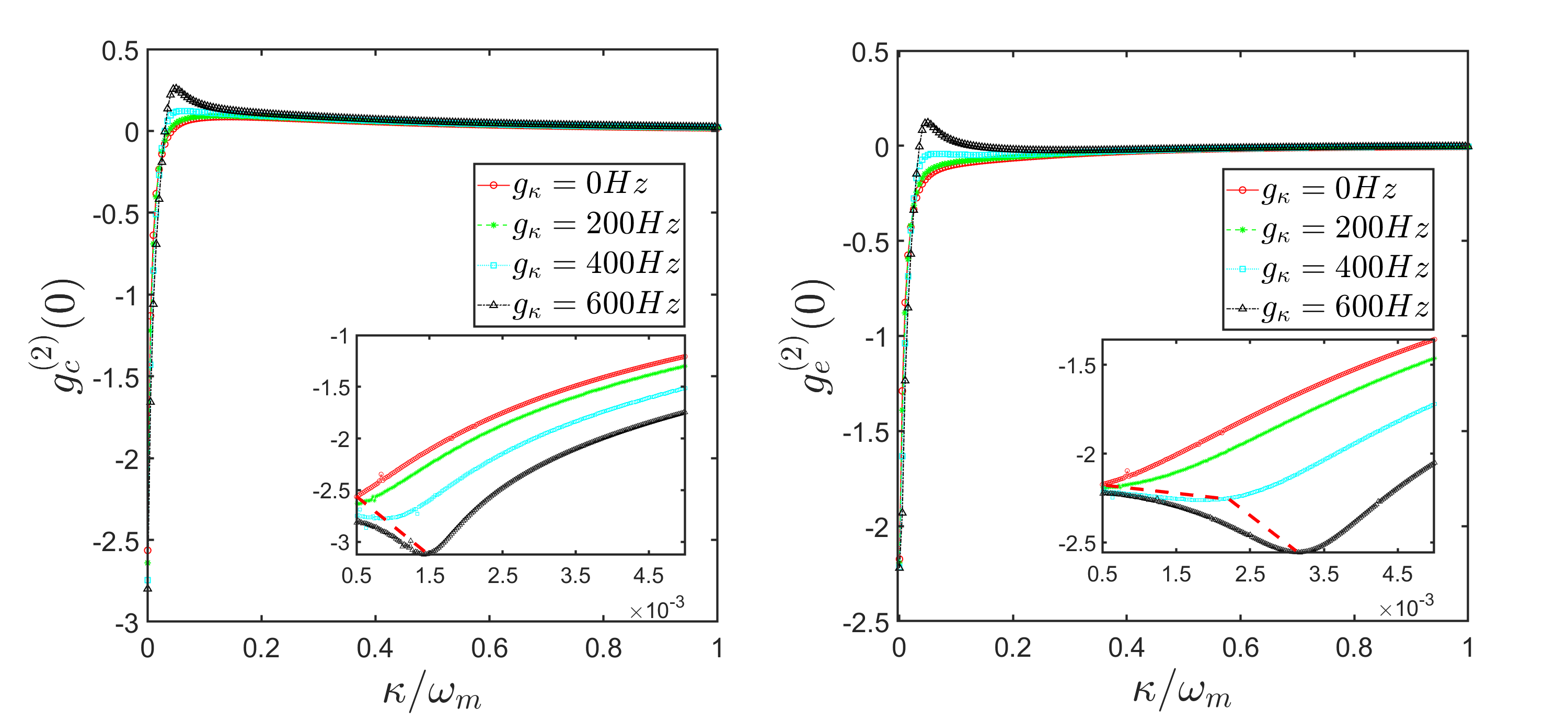}
    \caption{(Color online) Zero-time second-order correlation $g^2(0) $ as a function of the dissipation $\kappa/\omega_m$ in  two cavities. Where $g_\omega=200Hz$ does not change and $g_\kappa$ is 0, 200, 400 and 600 Hz, respectively. The other parameters are $\omega_m=10^6 Hz, \Delta_c=\Delta_e=-J=-2\times10^5 Hz, \epsilon_c=\epsilon_e=5\times10^3 Hz.$ }
    \label{fig:7}
\end{figure}

Besides the effect of detuning on the system, we still need to explore the contribution of dissipative and dispersive coupling to the second-order correlation function.
We select the detuning parameters of the near-blockade region in \fref{fig:4} and \ref{fig:5} with $\Delta_c=\Delta_e=-J$ to investigate this contribution and illustrate in \fref{fig:6}.

As shown in \fref{fig:6}, the second-order correlation function $g^{(2)}_c(0)$ is the complex nonlinear relationship with the parameters $g_\omega$ and $g_\kappa$  for both the OM- and empty-cavities. Obviously, a smaller $g_\omega$ requires lager  $g_\kappa$ to obtain photon blockade effect for both the OM- and empty-cavities,  as is marked in I-regions of \fref{fig:6}(a) and \fref{fig:6}(b).
The special regions (Region-II in \fref{fig:6})
 with $\log_{10}[g^{(2)}(0)] \le -3$ are also marked to observe the optimal $g_c^{(2)}(0)$.
Compared with the case of non- dissipative coupling (corresponding to the regions with $g_\kappa=0$ in the figure), an improved photon blockade is obtained with the help of dissipative coupling.
This conclusion is consistent with the trend in \fref{fig:3}, i.e., the minimum value of the numerical solution in (b) is smaller than that in (a) at $\Delta=-J$.
\par The impact of dissipation on the system is shown in \fref{fig:7}. 
The second-order correlation function $g_c^{(2)}(0)$ will reach the minimum values as $\kappa$ increases. In addition, the  minimal values decrease as the dissipative coupling strength $g_\kappa$ increases, which is described by  the red- dotted line in the graph. 
This phenomenon can be explained by Eq. (\ref{eq.5}).   
When the dissipative coupling term $g_\kappa g_\omega /2\omega_m \hat{a}^\dag_c \hat{a}_c \hat{a}_c$ is equal to the optical dissipative term $\kappa_c/2 \hat{a}_c$, the system can be treated as a dissipation-free system. Then the nonlinear effect of the system can be strengthened, thus the minimum values of the second-order correlation function $g_c^{(2)}(0)$ will increase.
This process is similar to the inverse dissipation in non-Hermitian Hamiltonian and is discussed in the analysis of \eref{eq.7}.
The minimum second-order correlation function is derived as $\kappa_{min} \propto \frac{g_\kappa g_\omega}{2\omega_m}$.
Obviously, $\kappa_{min}$ is increased with the rising of  $g_\kappa g_\omega$.

\begin{figure}[htp!]
    \centering
  \includegraphics[trim=0 20 0 20,clip,width=8.5cm]{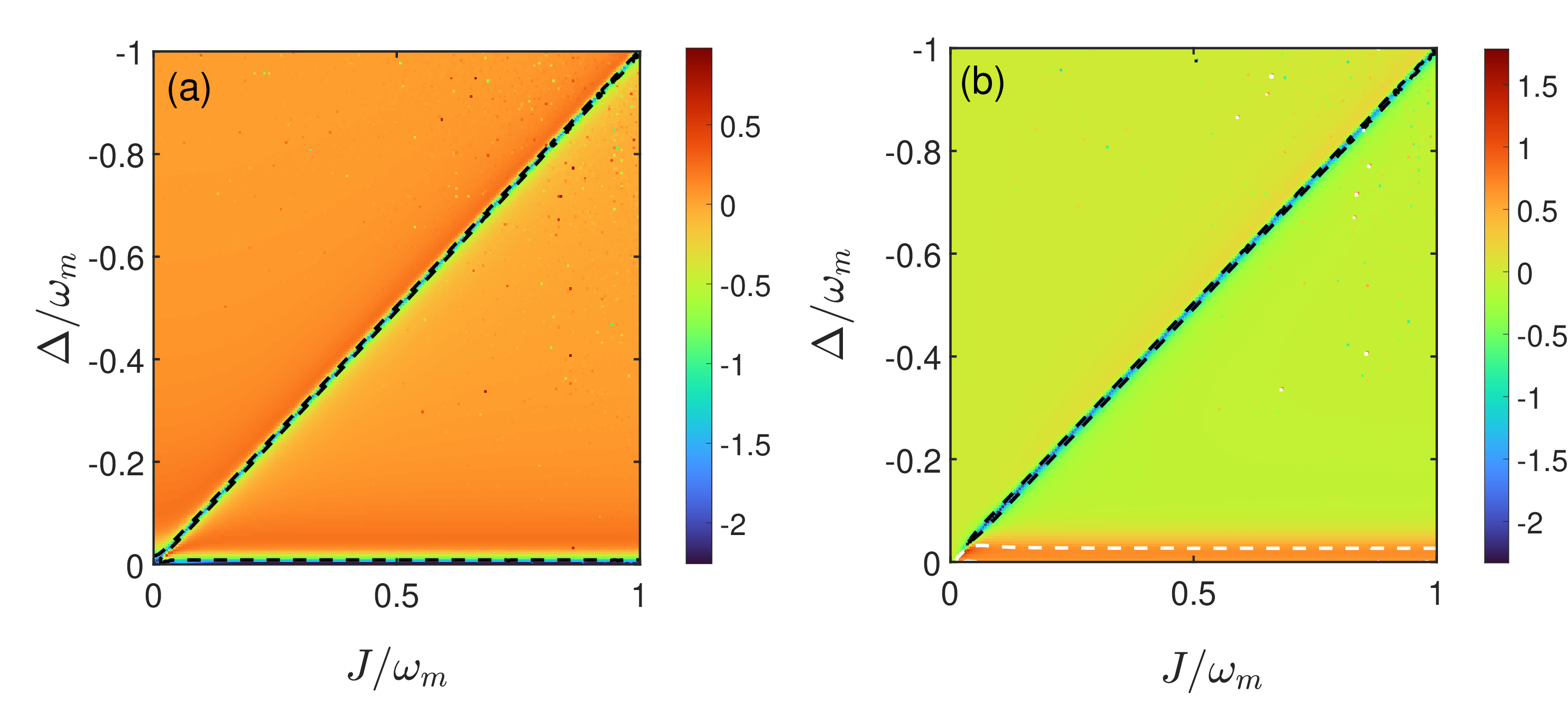}
    \caption{(Color online) Zero-time second-order correlation $g^{(2)}(0) $ as a function of the cavity coupling strength J and detuning $\Delta$ in optomechanical cavity (a) and auxiliary cavity (b).The black dashed line represents $log_{10}[g^2(0)] = -1$ and the white dashed line represents $log_{10}[g^2(0)] = 0.602$. The other parameters are $\omega_m=10^6 Hz, \kappa=5\times10^3 Hz, \Delta_c=\Delta_e=\Delta
,g_\omega=200Hz,g_\kappa=500Hz,\epsilon_c=\epsilon_e=5\times10^3 Hz.$ }
    \label{fig:8}
\end{figure}
\par According to the previous analysis, $J$ is  a very important parameter which can be tuned in waveguide-coupled and close-coupled optical systems \cite{OE.18.014926,OLE.127.105968}.
As shown in \fref{fig:8}, the effect of $J$ and the detuning on the second-order correlation function  $g^{(2)}(0) $ is analysed.
 We can see that the photon blockade effect (black-dashed line) only occurs when $J = -\Delta$. While the second-order correlation function  $g^{(2)}(0) $ increases rapidly when $J\neq -\Delta$. Moreover,
optomechanical cavity shows anti-bunching effect while auxiliary cavity shows bunching effect (white-dashed line) for $\Delta=0$.
This difference can be understood by the analytical solution. 
In \eref{eq.11}, we have derived $g_c^{(2)}(0)=0$ and  $g_e^{(2)}(0)=4$ at $\Delta = 0$.

\section{Discussion and Conclusion}\label{sec4}
In our discussion, we have chosen the eigenfrequency of the mechanical oscillator in the order of $10^6$Hz, which is realized  in experiments for optomechanical systems \cite{Ncomms.6.8491,PhysRevLett.97.133601,RevModPhys.86.1391}. Recently, it was experimentally reported 
the frequency of SiN membrane in the MSI-based dissipative coupled systems is $\omega_m=2\pi\times 136$kHz \cite{PhysRevLett.114.043601}.
According to the detailed discussion in \fref{fig:7},  we can obtain $g^{(2)}_j(0)<10^{-1}$ with $\kappa/\omega_m<10^{-2}$. 
This requirement under resolved sideband has been realised in some optomechanical systems \cite{nature.475.359,nphys.5.489,PhysRevLett.126.123603}. At present,
for single-photon coupling rates, it is still in the weak coupling regime in experiments \cite{nature.471.204,Nature.488.476,Ncomms.6.8491,PhysRevLett.126.123603}, i.e. $g/\omega_m\in \{10^{-9}-10^{-1} \}$. 
In \fref{fig:4} and \ref{fig:5}, we show that it is possible to realise strong anti-bunching effect in the weak coupling regime with $g_{\omega(\kappa)}/\omega_m$ in the order $10^{-4}-10^{-3}$.
In addition, the requirement of the temperature in our system is compatible to the related experiments. The effective environmental temperature is 7 to 8 orders of magnitude smaller than the real temperature  (see \fref{fig:2}).

In conclusion,  a full-quantum method is applied  to study the dissipative coupled system without using linearized approximation. 
A strong anti-bunching effect is realized in nonlinear weak coupling regime.
The optimal region of second-order correlation function is a hyperbolic form with relationship $J^2 = \Delta_c \Delta_e$. Moreover, the system exhibit photon blockade effect with the appropriate parameters for $g_\omega$ and $g_\kappa$. Especially, the dissipative coupling in our system  can resist the destruction of the quantum effects by conventional dissipation.

Going forward, we confirm the validity of the dissipative coupling theory to study the quantum nonlinear effects in optomechanical systems.
Our quantum nonlinear effects require neither strong single-photon coupling coefficients as in conventional blockade \cite{PhysRevLett.107.063601,PhysRevA.88.023853,PhysRevA.100.063817} nor stringent parametrics condition as in unconventional blockade \cite{PhysRevA.96.053810,PhysRevA.90.063824,PhysRevA.98.013826,PhysRevA.98.023856}. 
This dissipative coupling-assisted blockade effect is very favorable for performing single-photon transmission control \cite{PhysRevA.91.063836} and implementing optical quantum diodes \cite{PhysRevLett.110.093901,PhysRevLett.94.067401}.
In quantum information processing, MSI-based optomechanical systems can also serve as a core to realize optical and mechanical-optical controlled gates \cite{PhysRevA.79.022301,JPB.48.015502,OE.23.7786}.
Moreover, this strong nonlinearity is a necessary resource in quantum synchronization \cite{PhysRevLett.111.103605,PhysRevE.95.022204,PhysRevLett.112.094102,PhysRevLett.109.233906} and the study of quantum chaos effects \cite{PhysRevLett.114.253601}.
Furthermore, MSI can serve as a kindly platform for exploring classical and quantum transitions in the nonlinear regime \cite{PhysRevLett.125.014101,nature.461.768}.

\section{Acknowledgments}
We appreciate Qiu Hui-Hui, Rui-Jie Xiao and Leng Xuan's constructive discussions. This work was partly supported by the National Natural Science Foundation of China under Grant Nos. 12074206 and 12265022, the Natural Science Foundation of Zhejiang Province under Grant No. LY22A040005, the Inner Mongolia Natural Science Foundation under Grant No. 2021MS01012, the Inner Mongolia Fundamental Research Funds for the 
directly affiliated Universities Grant No. 
2023RCTD014 and K. C. Wong Magna Fund in Ningbo University.

\appendix
\section{Approximate derivation of dissipative coupling} \label{app.1}
The cavity decay rate $\kappa_c(L+\hat{x})$ depends on the displacement of the mechanical oscillator $\hat{x}$. Since the displacement of the mechanical oscillator is very small compared to the cavity length, a Taylor expansion can be made for $\kappa_c(L+\hat{x})$:
\begin{eqnarray}
\kappa_c(L+\hat{x})&=&\kappa_c(L)+\frac{\partial \kappa_c(L)}{\partial L}\hat{x}+ \frac{\partial^2 \kappa_c(L)}{2\partial^2 L}\hat{x}^2 +o([\frac{\hat{x}}{L}]^3),\nonumber\\
\label{eq.A1}
\end{eqnarray}
with a movable ideal end mirror
and an input coupler of transmissivity $\tau$, the cavity linewidth $\kappa_c$ is given by \cite{PhysRevLett.107.213604}
\begin{eqnarray}
 \kappa_c(L+\hat{x})=\frac{c|\tau|^2}{4(L+\hat{x})}. 
 \label{eq.A2}
\end{eqnarray}
Substituting Eq. \ref{eq.A2} back to Eq. \ref{eq.A1} and ignoring the higher order terms, we can approximate to the first order term
\begin{eqnarray}
 \kappa_c(L+\hat{x})=\frac{c|\tau^2|}{4L}(1-\frac{\hat{x}}{L} ), 
\end{eqnarray}
with $\kappa_c=\kappa_c(L)=\frac{c|\tau|^2}{4L}$, $g_\kappa=\frac{\partial \kappa_c(L)}{\partial L}x_{zpf}=-\frac{\kappa_c}{L}x_{zpf}$. We derive   $L=-\frac{\kappa_c}{g_\kappa}x_{zpf}$ and the square root form with $\hat{x}=\hat{Q}x_{zpf}$:
\begin{eqnarray}
\sqrt{\kappa_c(L+\hat{x})}&=&\sqrt{\kappa _c+g_\kappa \hat{Q}}=\sqrt{\frac{c|\tau^2|}{4L}(1-\frac{\hat{x}}{L} )}\nonumber\\& \approx &\sqrt{\kappa _c}(1-\frac{\hat{x}}{2L})=\sqrt{\kappa _c}(1+\frac{g_\kappa}{2\kappa_c}\hat{Q}).\nonumber\\
\end{eqnarray}
\section{Derivation of second order correlation function} \label{app.2}
With the definition $D_J=J^2-\Delta_c\Delta_e$, $K=\Delta_c+\Delta_e$, $G=2g_\omega^2-i g_\kappa g_\omega$ and set $\kappa_c=\kappa_e=0$, $\epsilon_c=\epsilon_e=\epsilon$ under the weak-driving condition ($C_0\gg C_m\gg C_{mn}$), we can obtain the dynamic equations of the probability amplitudes. 
\begin{eqnarray}
i\dot{C}_c&=&\epsilon_c C_0+JC_e-\Delta_c C_c,\nonumber\\
i\dot{C}_e&=&\epsilon_e C_0+JC_c-\Delta_e C_e,\nonumber\\
i\dot{C}_{ce}&=&\epsilon_e C_c+\epsilon_c C_e+\sqrt{2}J(C_{ee}+C_{cc})-K C_{ce},\nonumber\\
i\dot{C}_{cc}&=&\sqrt{2}\epsilon_c C_c+\sqrt{2}JC_{ce}-2(\Delta_c+\frac{G}{2\omega_m}) C_{cc},\nonumber\\
i\dot{C}_{ee}&=&\sqrt{2}\epsilon_e C_e+\sqrt{2}JC_{ce}-2\Delta_e C_{ee}.
\end{eqnarray}
The corresponding steady-state solutions are obtained as 
\begin{eqnarray}
C_c&=&C_e=-\frac{\epsilon(J+\Delta_e)}{D_J}, \\  
C_{ce}&=&\epsilon^2\frac{[G(J+K)+2K\omega_m(J+\Delta_c)](J+\Delta_e)}{D_J[2KD_J\omega_m+G(J^2-K\Delta_e)]},    \nonumber\\
C_{cc}&=&\frac{\sqrt{2}K\epsilon^2\omega_m(J+\Delta_e)^2}{D_J[2KD_J\omega_m+G(J^2-K\Delta_e)]},    \nonumber\\
C_{ee}&=&\epsilon^2\frac{G[2J(J+K)+K\Delta_c]+2K\omega_m(J+\Delta_c)^2}{\sqrt{2}D_J[2KD_J\omega_m+G(J^2-K\Delta_e)]}.\nonumber
\end{eqnarray}
Thus, the second-order correlation functions with zero time delay are read:
\begin{eqnarray}
 g^{(2)}(0)&=&\frac{\sum^{2}_{n=1}n(n-1)P(n)}{[\sum^{2}_{n=1}n P(n)]^2},\\
 g^{(2)}_c(0)&=&\frac{2|C_{cc}|^2}{(|C_c|^2+|C_{ce}|^2+2|C_{cc}|^2)^2}\approx
 \frac{2|C_{cc}|^2}{|C_c|^4},\nonumber\\
g^{(2)}_e(0)&=&\frac{2|C_{ee}|^2}{(|C_e|^2+|C_{ce}|^2+2|C_{ee}|^2)^2}\approx
 \frac{2|C_{ee}|^2}{|C_e|^4}.\nonumber
\end{eqnarray}

\bibliography{dc}

\begin{thebibliography}{75}%
\makeatletter
\providecommand \@ifxundefined [1]{%
 \@ifx{#1\undefined}
}%
\providecommand \@ifnum [1]{%
 \ifnum #1\expandafter \@firstoftwo
 \else \expandafter \@secondoftwo
 \fi
}%
\providecommand \@ifx [1]{%
 \ifx #1\expandafter \@firstoftwo
 \else \expandafter \@secondoftwo
 \fi
}%
\providecommand \natexlab [1]{#1}%
\providecommand \enquote  [1]{``#1''}%
\providecommand \bibnamefont  [1]{#1}%
\providecommand \bibfnamefont [1]{#1}%
\providecommand \citenamefont [1]{#1}%
\providecommand \href@noop [0]{\@secondoftwo}%
\providecommand \href [0]{\begingroup \@sanitize@url \@href}%
\providecommand \@href[1]{\@@startlink{#1}\@@href}%
\providecommand \@@href[1]{\endgroup#1\@@endlink}%
\providecommand \@sanitize@url [0]{\catcode `\\12\catcode `\$12\catcode
  `\&12\catcode `\#12\catcode `\^12\catcode `\_12\catcode `\%12\relax}%
\providecommand \@@startlink[1]{}%
\providecommand \@@endlink[0]{}%
\providecommand \url  [0]{\begingroup\@sanitize@url \@url }%
\providecommand \@url [1]{\endgroup\@href {#1}{\urlprefix }}%
\providecommand \urlprefix  [0]{URL }%
\providecommand \Eprint [0]{\href }%
\providecommand \doibase [0]{http://dx.doi.org/}%
\providecommand \selectlanguage [0]{\@gobble}%
\providecommand \bibinfo  [0]{\@secondoftwo}%
\providecommand \bibfield  [0]{\@secondoftwo}%
\providecommand \translation [1]{[#1]}%
\providecommand \BibitemOpen [0]{}%
\providecommand \bibitemStop [0]{}%
\providecommand \bibitemNoStop [0]{.\EOS\space}%
\providecommand \EOS [0]{\spacefactor3000\relax}%
\providecommand \BibitemShut  [1]{\csname bibitem#1\endcsname}%
\let\auto@bib@innerbib\@empty
\bibitem [{\citenamefont {Aspelmeyer}\ \emph {et~al.}(2014)\citenamefont
  {Aspelmeyer}, \citenamefont {Kippenberg},\ and\ \citenamefont
  {Marquardt}}]{RevModPhys.86.1391}%
  \BibitemOpen
  \bibfield  {author} {\bibinfo {author} {\bibfnamefont {M.}~\bibnamefont
  {Aspelmeyer}}, \bibinfo {author} {\bibfnamefont {T.~J.}\ \bibnamefont
  {Kippenberg}}, \ and\ \bibinfo {author} {\bibfnamefont {F.}~\bibnamefont
  {Marquardt}},\ }\href {\doibase 10.1103/RevModPhys.86.1391} {\bibfield
  {journal} {\bibinfo  {journal} {Rev. Mod. Phys.}\ }\textbf {\bibinfo {volume}
  {86}},\ \bibinfo {pages} {1391} (\bibinfo {year} {2014})}\BibitemShut
  {NoStop}%
\bibitem [{\citenamefont {Kippenberg}\ and\ \citenamefont
  {Vahala}(2007)}]{Kippenberg:07}%
  \BibitemOpen
  \bibfield  {author} {\bibinfo {author} {\bibfnamefont {T.}~\bibnamefont
  {Kippenberg}}\ and\ \bibinfo {author} {\bibfnamefont {K.}~\bibnamefont
  {Vahala}},\ }\href {\doibase 10.1364/OE.15.017172} {\bibfield  {journal}
  {\bibinfo  {journal} {Opt. Express}\ }\textbf {\bibinfo {volume} {15}},\
  \bibinfo {pages} {17172} (\bibinfo {year} {2007})}\BibitemShut {NoStop}%
\bibitem [{\citenamefont {Kippenberg}\ and\ \citenamefont
  {Vahala}(2008)}]{doi:10.1126/science.1156032}%
  \BibitemOpen
  \bibfield  {author} {\bibinfo {author} {\bibfnamefont {T.~J.}\ \bibnamefont
  {Kippenberg}}\ and\ \bibinfo {author} {\bibfnamefont {K.~J.}\ \bibnamefont
  {Vahala}},\ }\href {\doibase 10.1126/science.1156032} {\bibfield  {journal}
  {\bibinfo  {journal} {Science}\ }\textbf {\bibinfo {volume} {321}},\ \bibinfo
  {pages} {1172} (\bibinfo {year} {2008})}\BibitemShut {NoStop}%
\bibitem [{\citenamefont {Faust}\ \emph {et~al.}(2012)\citenamefont {Faust},
  \citenamefont {Krenn}, \citenamefont {Manus}, \citenamefont {Kotthaus},\ and\
  \citenamefont {Weig}}]{faust2012microwave}%
  \BibitemOpen
  \bibfield  {author} {\bibinfo {author} {\bibfnamefont {T.}~\bibnamefont
  {Faust}}, \bibinfo {author} {\bibfnamefont {P.}~\bibnamefont {Krenn}},
  \bibinfo {author} {\bibfnamefont {S.}~\bibnamefont {Manus}}, \bibinfo
  {author} {\bibfnamefont {J.~P.}\ \bibnamefont {Kotthaus}}, \ and\ \bibinfo
  {author} {\bibfnamefont {E.~M.}\ \bibnamefont {Weig}},\ }\href@noop {}
  {\bibfield  {journal} {\bibinfo  {journal} {Nature communications}\ }\textbf
  {\bibinfo {volume} {3}},\ \bibinfo {pages} {728} (\bibinfo {year}
  {2012})}\BibitemShut {NoStop}%
\bibitem [{\citenamefont {Purdy}\ \emph {et~al.}(2013)\citenamefont {Purdy},
  \citenamefont {Yu}, \citenamefont {Peterson}, \citenamefont {Kampel},\ and\
  \citenamefont {Regal}}]{PhysRevX.3.031012}%
  \BibitemOpen
  \bibfield  {author} {\bibinfo {author} {\bibfnamefont {T.~P.}\ \bibnamefont
  {Purdy}}, \bibinfo {author} {\bibfnamefont {P.-L.}\ \bibnamefont {Yu}},
  \bibinfo {author} {\bibfnamefont {R.~W.}\ \bibnamefont {Peterson}}, \bibinfo
  {author} {\bibfnamefont {N.~S.}\ \bibnamefont {Kampel}}, \ and\ \bibinfo
  {author} {\bibfnamefont {C.~A.}\ \bibnamefont {Regal}},\ }\href {\doibase
  10.1103/PhysRevX.3.031012} {\bibfield  {journal} {\bibinfo  {journal} {Phys.
  Rev. X}\ }\textbf {\bibinfo {volume} {3}},\ \bibinfo {pages} {031012}
  (\bibinfo {year} {2013})}\BibitemShut {NoStop}%
\bibitem [{\citenamefont {Schnabel}(2017)}]{SCHNABEL20171}%
  \BibitemOpen
  \bibfield  {author} {\bibinfo {author} {\bibfnamefont {R.}~\bibnamefont
  {Schnabel}},\ }\href {\doibase https://doi.org/10.1016/j.physrep.2017.04.001}
  {\bibfield  {journal} {\bibinfo  {journal} {Physics Reports}\ }\textbf
  {\bibinfo {volume} {684}},\ \bibinfo {pages} {1} (\bibinfo {year} {2017})},\
  \bibinfo {note} {squeezed states of light and their applications in laser
  interferometers}\BibitemShut {NoStop}%
\bibitem [{\citenamefont {Paternostro}\ \emph {et~al.}(2007)\citenamefont
  {Paternostro}, \citenamefont {Vitali}, \citenamefont {Gigan}, \citenamefont
  {Kim}, \citenamefont {Brukner}, \citenamefont {Eisert},\ and\ \citenamefont
  {Aspelmeyer}}]{PhysRevLett.99.250401}%
  \BibitemOpen
  \bibfield  {author} {\bibinfo {author} {\bibfnamefont {M.}~\bibnamefont
  {Paternostro}}, \bibinfo {author} {\bibfnamefont {D.}~\bibnamefont {Vitali}},
  \bibinfo {author} {\bibfnamefont {S.}~\bibnamefont {Gigan}}, \bibinfo
  {author} {\bibfnamefont {M.~S.}\ \bibnamefont {Kim}}, \bibinfo {author}
  {\bibfnamefont {C.}~\bibnamefont {Brukner}}, \bibinfo {author} {\bibfnamefont
  {J.}~\bibnamefont {Eisert}}, \ and\ \bibinfo {author} {\bibfnamefont
  {M.}~\bibnamefont {Aspelmeyer}},\ }\href {\doibase
  10.1103/PhysRevLett.99.250401} {\bibfield  {journal} {\bibinfo  {journal}
  {Phys. Rev. Lett.}\ }\textbf {\bibinfo {volume} {99}},\ \bibinfo {pages}
  {250401} (\bibinfo {year} {2007})}\BibitemShut {NoStop}%
\bibitem [{\citenamefont {Lai}\ \emph {et~al.}(2022)\citenamefont {Lai},
  \citenamefont {Chen}, \citenamefont {Qin}, \citenamefont {Miranowicz},\ and\
  \citenamefont {Nori}}]{PhysRevResearch.4.033112}%
  \BibitemOpen
  \bibfield  {author} {\bibinfo {author} {\bibfnamefont {D.-G.}\ \bibnamefont
  {Lai}}, \bibinfo {author} {\bibfnamefont {Y.-H.}\ \bibnamefont {Chen}},
  \bibinfo {author} {\bibfnamefont {W.}~\bibnamefont {Qin}}, \bibinfo {author}
  {\bibfnamefont {A.}~\bibnamefont {Miranowicz}}, \ and\ \bibinfo {author}
  {\bibfnamefont {F.}~\bibnamefont {Nori}},\ }\href {\doibase
  10.1103/PhysRevResearch.4.033112} {\bibfield  {journal} {\bibinfo  {journal}
  {Phys. Rev. Res.}\ }\textbf {\bibinfo {volume} {4}},\ \bibinfo {pages}
  {033112} (\bibinfo {year} {2022})}\BibitemShut {NoStop}%
\bibitem [{\citenamefont {Bhattacharya}\ and\ \citenamefont
  {Meystre}(2007)}]{PhysRevLett.99.073601}%
  \BibitemOpen
  \bibfield  {author} {\bibinfo {author} {\bibfnamefont {M.}~\bibnamefont
  {Bhattacharya}}\ and\ \bibinfo {author} {\bibfnamefont {P.}~\bibnamefont
  {Meystre}},\ }\href {\doibase 10.1103/PhysRevLett.99.073601} {\bibfield
  {journal} {\bibinfo  {journal} {Phys. Rev. Lett.}\ }\textbf {\bibinfo
  {volume} {99}},\ \bibinfo {pages} {073601} (\bibinfo {year}
  {2007})}\BibitemShut {NoStop}%
\bibitem [{\citenamefont {Sawadsky}\ \emph {et~al.}(2015)\citenamefont
  {Sawadsky}, \citenamefont {Kaufer}, \citenamefont {Nia}, \citenamefont
  {Tarabrin}, \citenamefont {Khalili}, \citenamefont {Hammerer},\ and\
  \citenamefont {Schnabel}}]{PhysRevLett.114.043601}%
  \BibitemOpen
  \bibfield  {author} {\bibinfo {author} {\bibfnamefont {A.}~\bibnamefont
  {Sawadsky}}, \bibinfo {author} {\bibfnamefont {H.}~\bibnamefont {Kaufer}},
  \bibinfo {author} {\bibfnamefont {R.~M.}\ \bibnamefont {Nia}}, \bibinfo
  {author} {\bibfnamefont {S.~P.}\ \bibnamefont {Tarabrin}}, \bibinfo {author}
  {\bibfnamefont {F.~Y.}\ \bibnamefont {Khalili}}, \bibinfo {author}
  {\bibfnamefont {K.}~\bibnamefont {Hammerer}}, \ and\ \bibinfo {author}
  {\bibfnamefont {R.}~\bibnamefont {Schnabel}},\ }\href {\doibase
  10.1103/PhysRevLett.114.043601} {\bibfield  {journal} {\bibinfo  {journal}
  {Phys. Rev. Lett.}\ }\textbf {\bibinfo {volume} {114}},\ \bibinfo {pages}
  {043601} (\bibinfo {year} {2015})}\BibitemShut {NoStop}%
\bibitem [{\citenamefont {Huang}\ and\ \citenamefont
  {Agarwal}(2010)}]{PhysRevA.81.053810}%
  \BibitemOpen
  \bibfield  {author} {\bibinfo {author} {\bibfnamefont {S.}~\bibnamefont
  {Huang}}\ and\ \bibinfo {author} {\bibfnamefont {G.~S.}\ \bibnamefont
  {Agarwal}},\ }\href {\doibase 10.1103/PhysRevA.81.053810} {\bibfield
  {journal} {\bibinfo  {journal} {Phys. Rev. A}\ }\textbf {\bibinfo {volume}
  {81}},\ \bibinfo {pages} {053810} (\bibinfo {year} {2010})}\BibitemShut
  {NoStop}%
\bibitem [{\citenamefont {Weiss}\ \emph {et~al.}(2013)\citenamefont {Weiss},
  \citenamefont {Bruder},\ and\ \citenamefont {Nunnenkamp}}]{Weiss_2013}%
  \BibitemOpen
  \bibfield  {author} {\bibinfo {author} {\bibfnamefont {T.}~\bibnamefont
  {Weiss}}, \bibinfo {author} {\bibfnamefont {C.}~\bibnamefont {Bruder}}, \
  and\ \bibinfo {author} {\bibfnamefont {A.}~\bibnamefont {Nunnenkamp}},\
  }\href {\doibase 10.1088/1367-2630/15/4/045017} {\bibfield  {journal}
  {\bibinfo  {journal} {New Journal of Physics}\ }\textbf {\bibinfo {volume}
  {15}},\ \bibinfo {pages} {045017} (\bibinfo {year} {2013})}\BibitemShut
  {NoStop}%
\bibitem [{\citenamefont {Rabl}(2011)}]{PhysRevLett.107.063601}%
  \BibitemOpen
  \bibfield  {author} {\bibinfo {author} {\bibfnamefont {P.}~\bibnamefont
  {Rabl}},\ }\href {\doibase 10.1103/PhysRevLett.107.063601} {\bibfield
  {journal} {\bibinfo  {journal} {Phys. Rev. Lett.}\ }\textbf {\bibinfo
  {volume} {107}},\ \bibinfo {pages} {063601} (\bibinfo {year}
  {2011})}\BibitemShut {NoStop}%
\bibitem [{\citenamefont {Nunnenkamp}\ \emph {et~al.}(2011)\citenamefont
  {Nunnenkamp}, \citenamefont {B\o{}rkje},\ and\ \citenamefont
  {Girvin}}]{PhysRevLett.107.063602}%
  \BibitemOpen
  \bibfield  {author} {\bibinfo {author} {\bibfnamefont {A.}~\bibnamefont
  {Nunnenkamp}}, \bibinfo {author} {\bibfnamefont {K.}~\bibnamefont
  {B\o{}rkje}}, \ and\ \bibinfo {author} {\bibfnamefont {S.~M.}\ \bibnamefont
  {Girvin}},\ }\href {\doibase 10.1103/PhysRevLett.107.063602} {\bibfield
  {journal} {\bibinfo  {journal} {Phys. Rev. Lett.}\ }\textbf {\bibinfo
  {volume} {107}},\ \bibinfo {pages} {063602} (\bibinfo {year}
  {2011})}\BibitemShut {NoStop}%
\bibitem [{\citenamefont {Liew}\ and\ \citenamefont
  {Savona}(2010)}]{PhysRevLett.104.183601}%
  \BibitemOpen
  \bibfield  {author} {\bibinfo {author} {\bibfnamefont {T.~C.~H.}\
  \bibnamefont {Liew}}\ and\ \bibinfo {author} {\bibfnamefont {V.}~\bibnamefont
  {Savona}},\ }\href {\doibase 10.1103/PhysRevLett.104.183601} {\bibfield
  {journal} {\bibinfo  {journal} {Phys. Rev. Lett.}\ }\textbf {\bibinfo
  {volume} {104}},\ \bibinfo {pages} {183601} (\bibinfo {year}
  {2010})}\BibitemShut {NoStop}%
\bibitem [{\citenamefont {Miranowicz}\ \emph {et~al.}(2013)\citenamefont
  {Miranowicz}, \citenamefont {Paprzycka}, \citenamefont {Liu}, \citenamefont
  {Bajer},\ and\ \citenamefont {Nori}}]{PhysRevA.87.023809}%
  \BibitemOpen
  \bibfield  {author} {\bibinfo {author} {\bibfnamefont {A.}~\bibnamefont
  {Miranowicz}}, \bibinfo {author} {\bibfnamefont {M.}~\bibnamefont
  {Paprzycka}}, \bibinfo {author} {\bibfnamefont {Y.X.}\ \bibnamefont {Liu}},
  \bibinfo {author} {\bibfnamefont {J.}\ \bibnamefont {Bajer}}, \ and\
  \bibinfo {author} {\bibfnamefont {F.}~\bibnamefont {Nori}},\ }\href {\doibase
  10.1103/PhysRevA.87.023809} {\bibfield  {journal} {\bibinfo  {journal} {Phys.
  Rev. A}\ }\textbf {\bibinfo {volume} {87}},\ \bibinfo {pages} {023809}
  (\bibinfo {year} {2013})}\BibitemShut {NoStop}%
\bibitem [{\citenamefont {Huang}\ \emph {et~al.}(2018)\citenamefont {Huang},
  \citenamefont {Miranowicz}, \citenamefont {Liao}, \citenamefont {Nori},\ and\
  \citenamefont {Jing}}]{PhysRevLett.121.153601}%
  \BibitemOpen
  \bibfield  {author} {\bibinfo {author} {\bibfnamefont {R.}~\bibnamefont
  {Huang}}, \bibinfo {author} {\bibfnamefont {A.}~\bibnamefont {Miranowicz}},
  \bibinfo {author} {\bibfnamefont {J.-Q.}\ \bibnamefont {Liao}}, \bibinfo
  {author} {\bibfnamefont {F.}~\bibnamefont {Nori}}, \ and\ \bibinfo {author}
  {\bibfnamefont {H.}~\bibnamefont {Jing}},\ }\href {\doibase
  10.1103/PhysRevLett.121.153601} {\bibfield  {journal} {\bibinfo  {journal}
  {Phys. Rev. Lett.}\ }\textbf {\bibinfo {volume} {121}},\ \bibinfo {pages}
  {153601} (\bibinfo {year} {2018})}\BibitemShut {NoStop}%
\bibitem [{\citenamefont {Aldana}\ \emph {et~al.}(2013)\citenamefont {Aldana},
  \citenamefont {Bruder},\ and\ \citenamefont
  {Nunnenkamp}}]{PhysRevA.88.043826}%
  \BibitemOpen
  \bibfield  {author} {\bibinfo {author} {\bibfnamefont {S.}~\bibnamefont
  {Aldana}}, \bibinfo {author} {\bibfnamefont {C.}~\bibnamefont {Bruder}}, \
  and\ \bibinfo {author} {\bibfnamefont {A.}~\bibnamefont {Nunnenkamp}},\
  }\href {\doibase 10.1103/PhysRevA.88.043826} {\bibfield  {journal} {\bibinfo
  {journal} {Phys. Rev. A}\ }\textbf {\bibinfo {volume} {88}},\ \bibinfo
  {pages} {043826} (\bibinfo {year} {2013})}\BibitemShut {NoStop}%
\bibitem [{\citenamefont {Zhou}\ \emph {et~al.}(2013)\citenamefont {Zhou},
  \citenamefont {Cheng}, \citenamefont {Han},\ and\ \citenamefont
  {Zhang}}]{PhysRevA.88.063854}%
  \BibitemOpen
  \bibfield  {author} {\bibinfo {author} {\bibfnamefont {L.}~\bibnamefont
  {Zhou}}, \bibinfo {author} {\bibfnamefont {J.}~\bibnamefont {Cheng}},
  \bibinfo {author} {\bibfnamefont {Y.}~\bibnamefont {Han}}, \ and\ \bibinfo
  {author} {\bibfnamefont {W.}~\bibnamefont {Zhang}},\ }\href {\doibase
  10.1103/PhysRevA.88.063854} {\bibfield  {journal} {\bibinfo  {journal} {Phys.
  Rev. A}\ }\textbf {\bibinfo {volume} {88}},\ \bibinfo {pages} {063854}
  (\bibinfo {year} {2013})}\BibitemShut {NoStop}%
\bibitem [{\citenamefont {Weis}\ \emph {et~al.}(2010)\citenamefont {Weis},
  \citenamefont {Rivière}, \citenamefont {Deléglise}, \citenamefont
  {Gavartin}, \citenamefont {Arcizet}, \citenamefont {Schliesser},\ and\
  \citenamefont {Kippenberg}}]{doi:10.1126/science.1195596}%
  \BibitemOpen
  \bibfield  {author} {\bibinfo {author} {\bibfnamefont {S.}~\bibnamefont
  {Weis}}, \bibinfo {author} {\bibfnamefont {R.}~\bibnamefont {Rivière}},
  \bibinfo {author} {\bibfnamefont {S.}~\bibnamefont {Deléglise}}, \bibinfo
  {author} {\bibfnamefont {E.}~\bibnamefont {Gavartin}}, \bibinfo {author}
  {\bibfnamefont {O.}~\bibnamefont {Arcizet}}, \bibinfo {author} {\bibfnamefont
  {A.}~\bibnamefont {Schliesser}}, \ and\ \bibinfo {author} {\bibfnamefont
  {T.~J.}\ \bibnamefont {Kippenberg}},\ }\href {\doibase
  10.1126/science.1195596} {\bibfield  {journal} {\bibinfo  {journal}
  {Science}\ }\textbf {\bibinfo {volume} {330}},\ \bibinfo {pages} {1520}
  (\bibinfo {year} {2010})}\BibitemShut {NoStop}%
\bibitem [{\citenamefont {Kronwald}\ and\ \citenamefont
  {Marquardt}(2013)}]{PhysRevLett.111.133601}%
  \BibitemOpen
  \bibfield  {author} {\bibinfo {author} {\bibfnamefont {A.}~\bibnamefont
  {Kronwald}}\ and\ \bibinfo {author} {\bibfnamefont {F.}~\bibnamefont
  {Marquardt}},\ }\href {\doibase 10.1103/PhysRevLett.111.133601} {\bibfield
  {journal} {\bibinfo  {journal} {Phys. Rev. Lett.}\ }\textbf {\bibinfo
  {volume} {111}},\ \bibinfo {pages} {133601} (\bibinfo {year}
  {2013})}\BibitemShut {NoStop}%
\bibitem [{\citenamefont {Agarwal}\ and\ \citenamefont
  {Huang}(2010)}]{PhysRevA.81.041803}%
  \BibitemOpen
  \bibfield  {author} {\bibinfo {author} {\bibfnamefont {G.~S.}\ \bibnamefont
  {Agarwal}}\ and\ \bibinfo {author} {\bibfnamefont {S.}~\bibnamefont
  {Huang}},\ }\href {\doibase 10.1103/PhysRevA.81.041803} {\bibfield  {journal}
  {\bibinfo  {journal} {Phys. Rev. A}\ }\textbf {\bibinfo {volume} {81}},\
  \bibinfo {pages} {041803(R)} (\bibinfo {year} {2010})}\BibitemShut {NoStop}%
\bibitem [{\citenamefont {Teufel}\ \emph
  {et~al.}(2011{\natexlab{a}})\citenamefont {Teufel}, \citenamefont {Li},
  \citenamefont {Allman}, \citenamefont {Cicak}, \citenamefont {Sirois},
  \citenamefont {Whittaker},\ and\ \citenamefont
  {Simmonds}}]{teufel2011circuit}%
  \BibitemOpen
  \bibfield  {author} {\bibinfo {author} {\bibfnamefont {J.~D.}\ \bibnamefont
  {Teufel}}, \bibinfo {author} {\bibfnamefont {D.}~\bibnamefont {Li}}, \bibinfo
  {author} {\bibfnamefont {M.}~\bibnamefont {Allman}}, \bibinfo {author}
  {\bibfnamefont {K.}~\bibnamefont {Cicak}}, \bibinfo {author} {\bibfnamefont
  {A.}~\bibnamefont {Sirois}}, \bibinfo {author} {\bibfnamefont
  {J.}~\bibnamefont {Whittaker}}, \ and\ \bibinfo {author} {\bibfnamefont
  {R.}~\bibnamefont {Simmonds}},\ }\href {\doibase 10.1038/nature09898}
  {\bibfield  {journal} {\bibinfo  {journal} {Nature}\ }\textbf {\bibinfo
  {volume} {471}},\ \bibinfo {pages} {204} (\bibinfo {year}
  {2011}{\natexlab{a}})}\BibitemShut {NoStop}%
\bibitem [{\citenamefont {Law}(1995)}]{PhysRevA.51.2537}%
  \BibitemOpen
  \bibfield  {author} {\bibinfo {author} {\bibfnamefont {C.~K.}\ \bibnamefont
  {Law}},\ }\href {\doibase 10.1103/PhysRevA.51.2537} {\bibfield  {journal}
  {\bibinfo  {journal} {Phys. Rev. A}\ }\textbf {\bibinfo {volume} {51}},\
  \bibinfo {pages} {2537} (\bibinfo {year} {1995})}\BibitemShut {NoStop}%
\bibitem [{\citenamefont {Li}\ \emph {et~al.}(2009)\citenamefont {Li},
  \citenamefont {Pernice},\ and\ \citenamefont
  {Tang}}]{PhysRevLett.103.223901}%
  \BibitemOpen
  \bibfield  {author} {\bibinfo {author} {\bibfnamefont {M.}~\bibnamefont
  {Li}}, \bibinfo {author} {\bibfnamefont {W.~H.~P.}\ \bibnamefont {Pernice}},
  \ and\ \bibinfo {author} {\bibfnamefont {H.~X.}\ \bibnamefont {Tang}},\
  }\href {\doibase 10.1103/PhysRevLett.103.223901} {\bibfield  {journal}
  {\bibinfo  {journal} {Phys. Rev. Lett.}\ }\textbf {\bibinfo {volume} {103}},\
  \bibinfo {pages} {223901} (\bibinfo {year} {2009})}\BibitemShut {NoStop}%
\bibitem [{\citenamefont {Elste}\ \emph {et~al.}(2009)\citenamefont {Elste},
  \citenamefont {Girvin},\ and\ \citenamefont
  {Clerk}}]{PhysRevLett.102.207209}%
  \BibitemOpen
  \bibfield  {author} {\bibinfo {author} {\bibfnamefont {F.}~\bibnamefont
  {Elste}}, \bibinfo {author} {\bibfnamefont {S.~M.}\ \bibnamefont {Girvin}}, \
  and\ \bibinfo {author} {\bibfnamefont {A.~A.}\ \bibnamefont {Clerk}},\ }\href
  {\doibase 10.1103/PhysRevLett.102.207209} {\bibfield  {journal} {\bibinfo
  {journal} {Phys. Rev. Lett.}\ }\textbf {\bibinfo {volume} {102}},\ \bibinfo
  {pages} {207209} (\bibinfo {year} {2009})}\BibitemShut {NoStop}%
\bibitem [{\citenamefont {Xuereb}\ \emph {et~al.}(2011)\citenamefont {Xuereb},
  \citenamefont {Schnabel},\ and\ \citenamefont
  {Hammerer}}]{PhysRevLett.107.213604}%
  \BibitemOpen
  \bibfield  {author} {\bibinfo {author} {\bibfnamefont {A.}~\bibnamefont
  {Xuereb}}, \bibinfo {author} {\bibfnamefont {R.}~\bibnamefont {Schnabel}}, \
  and\ \bibinfo {author} {\bibfnamefont {K.}~\bibnamefont {Hammerer}},\ }\href
  {\doibase 10.1103/PhysRevLett.107.213604} {\bibfield  {journal} {\bibinfo
  {journal} {Phys. Rev. Lett.}\ }\textbf {\bibinfo {volume} {107}},\ \bibinfo
  {pages} {213604} (\bibinfo {year} {2011})}\BibitemShut {NoStop}%
\bibitem [{\citenamefont {Tagantsev}(2020)}]{PhysRevA.102.043520}%
  \BibitemOpen
  \bibfield  {author} {\bibinfo {author} {\bibfnamefont {A.~K.}\ \bibnamefont
  {Tagantsev}},\ }\href {\doibase 10.1103/PhysRevA.102.043520} {\bibfield
  {journal} {\bibinfo  {journal} {Phys. Rev. A}\ }\textbf {\bibinfo {volume}
  {102}},\ \bibinfo {pages} {043520} (\bibinfo {year} {2020})}\BibitemShut
  {NoStop}%
\bibitem [{\citenamefont {Huang}\ \emph {et~al.}(2023)\citenamefont {Huang},
  \citenamefont {Deng},\ and\ \citenamefont {Chen}}]{PhysRevA.107.013524}%
  \BibitemOpen
  \bibfield  {author} {\bibinfo {author} {\bibfnamefont {S.}~\bibnamefont
  {Huang}}, \bibinfo {author} {\bibfnamefont {L.}~\bibnamefont {Deng}}, \ and\
  \bibinfo {author} {\bibfnamefont {A.}~\bibnamefont {Chen}},\ }\href {\doibase
  10.1103/PhysRevA.107.013524} {\bibfield  {journal} {\bibinfo  {journal}
  {Phys. Rev. A}\ }\textbf {\bibinfo {volume} {107}},\ \bibinfo {pages}
  {013524} (\bibinfo {year} {2023})}\BibitemShut {NoStop}%
\bibitem [{\citenamefont {Qu}\ and\ \citenamefont
  {Agarwal}(2015)}]{PhysRevA.91.063815}%
  \BibitemOpen
  \bibfield  {author} {\bibinfo {author} {\bibfnamefont {K.}~\bibnamefont
  {Qu}}\ and\ \bibinfo {author} {\bibfnamefont {G.~S.}\ \bibnamefont
  {Agarwal}},\ }\href {\doibase 10.1103/PhysRevA.91.063815} {\bibfield
  {journal} {\bibinfo  {journal} {Phys. Rev. A}\ }\textbf {\bibinfo {volume}
  {91}},\ \bibinfo {pages} {063815} (\bibinfo {year} {2015})}\BibitemShut
  {NoStop}%
\bibitem [{\citenamefont {Kilda}\ and\ \citenamefont
  {Nunnenkamp}(2015)}]{Kilda_2016}%
  \BibitemOpen
  \bibfield  {author} {\bibinfo {author} {\bibfnamefont {D.}~\bibnamefont
  {Kilda}}\ and\ \bibinfo {author} {\bibfnamefont {A.}~\bibnamefont
  {Nunnenkamp}},\ }\href {\doibase 10.1088/2040-8978/18/1/014007} {\bibfield
  {journal} {\bibinfo  {journal} {Journal of Optics}\ }\textbf {\bibinfo
  {volume} {18}},\ \bibinfo {pages} {014007} (\bibinfo {year}
  {2015})}\BibitemShut {NoStop}%
\bibitem [{\citenamefont {Tagantsev}\ \emph {et~al.}(2018)\citenamefont
  {Tagantsev}, \citenamefont {Sokolov},\ and\ \citenamefont
  {Polzik}}]{PhysRevA.97.063820}%
  \BibitemOpen
  \bibfield  {author} {\bibinfo {author} {\bibfnamefont {A.~K.}\ \bibnamefont
  {Tagantsev}}, \bibinfo {author} {\bibfnamefont {I.~V.}\ \bibnamefont
  {Sokolov}}, \ and\ \bibinfo {author} {\bibfnamefont {E.~S.}\ \bibnamefont
  {Polzik}},\ }\href {\doibase 10.1103/PhysRevA.97.063820} {\bibfield
  {journal} {\bibinfo  {journal} {Phys. Rev. A}\ }\textbf {\bibinfo {volume}
  {97}},\ \bibinfo {pages} {063820} (\bibinfo {year} {2018})}\BibitemShut
  {NoStop}%
\bibitem [{\citenamefont {Meyer}\ \emph {et~al.}(2016)\citenamefont {Meyer},
  \citenamefont {Breyer},\ and\ \citenamefont
  {Köhl}}]{10.1007/s00340-016-6564-z}%
  \BibitemOpen
  \bibfield  {author} {\bibinfo {author} {\bibfnamefont {H.~M.}\ \bibnamefont
  {Meyer}}, \bibinfo {author} {\bibfnamefont {M.}~\bibnamefont {Breyer}}, \
  and\ \bibinfo {author} {\bibfnamefont {M.}~\bibnamefont {Köhl}},\ }\href
  {\doibase 10.1007/s00340-016-6564-z} {\bibfield  {journal} {\bibinfo
  {journal} {Applied Physics B}\ }\textbf {\bibinfo {volume} {122}},\ \bibinfo
  {pages} {1432} (\bibinfo {year} {2016})}\BibitemShut {NoStop}%
\bibitem [{\citenamefont {Madugani}\ \emph {et~al.}(2015)\citenamefont
  {Madugani}, \citenamefont {Yang}, \citenamefont {Ward}, \citenamefont {Le},\
  and\ \citenamefont {Nic~Chormaic}}]{10.1063/1.4922637}%
  \BibitemOpen
  \bibfield  {author} {\bibinfo {author} {\bibfnamefont {R.}~\bibnamefont
  {Madugani}}, \bibinfo {author} {\bibfnamefont {Y.~.}\ \bibnamefont {Yang}},
  \bibinfo {author} {\bibfnamefont {J.~M.}\ \bibnamefont {Ward}}, \bibinfo
  {author} {\bibfnamefont {V.~H.}\ \bibnamefont {Le}}, \ and\ \bibinfo {author}
  {\bibfnamefont {S.}~\bibnamefont {Nic~Chormaic}},\ }\href {\doibase
  10.1063/1.4922637} {\bibfield  {journal} {\bibinfo  {journal} {Applied
  Physics Letters}\ }\textbf {\bibinfo {volume} {106}},\ \bibinfo {pages}
  {241101} (\bibinfo {year} {2015})}\BibitemShut {NoStop}%
\bibitem [{\citenamefont {Meng}\ \emph {et~al.}(2022)\citenamefont {Meng},
  \citenamefont {Tang}, \citenamefont {Sun}, \citenamefont {Shen},
  \citenamefont {Li}, \citenamefont {Gong},\ and\ \citenamefont
  {Xiao}}]{PhysRevLett.129.073901}%
  \BibitemOpen
  \bibfield  {author} {\bibinfo {author} {\bibfnamefont {J.-W.}\ \bibnamefont
  {Meng}}, \bibinfo {author} {\bibfnamefont {S.-J.}\ \bibnamefont {Tang}},
  \bibinfo {author} {\bibfnamefont {J.}~\bibnamefont {Sun}}, \bibinfo {author}
  {\bibfnamefont {K.}~\bibnamefont {Shen}}, \bibinfo {author} {\bibfnamefont
  {C.}~\bibnamefont {Li}}, \bibinfo {author} {\bibfnamefont {Q.}~\bibnamefont
  {Gong}}, \ and\ \bibinfo {author} {\bibfnamefont {Y.-F.}\ \bibnamefont
  {Xiao}},\ }\href {\doibase 10.1103/PhysRevLett.129.073901} {\bibfield
  {journal} {\bibinfo  {journal} {Phys. Rev. Lett.}\ }\textbf {\bibinfo
  {volume} {129}},\ \bibinfo {pages} {073901} (\bibinfo {year}
  {2022})}\BibitemShut {NoStop}%
\bibitem [{\citenamefont {Wu}\ \emph {et~al.}(2014)\citenamefont {Wu},
  \citenamefont {Hryciw}, \citenamefont {Healey}, \citenamefont {Lake},
  \citenamefont {Jayakumar}, \citenamefont {Freeman}, \citenamefont {Davis},\
  and\ \citenamefont {Barclay}}]{PhysRevX.4.021052}%
  \BibitemOpen
  \bibfield  {author} {\bibinfo {author} {\bibfnamefont {M.}~\bibnamefont
  {Wu}}, \bibinfo {author} {\bibfnamefont {A.~C.}\ \bibnamefont {Hryciw}},
  \bibinfo {author} {\bibfnamefont {C.}~\bibnamefont {Healey}}, \bibinfo
  {author} {\bibfnamefont {D.~P.}\ \bibnamefont {Lake}}, \bibinfo {author}
  {\bibfnamefont {H.}~\bibnamefont {Jayakumar}}, \bibinfo {author}
  {\bibfnamefont {M.~R.}\ \bibnamefont {Freeman}}, \bibinfo {author}
  {\bibfnamefont {J.~P.}\ \bibnamefont {Davis}}, \ and\ \bibinfo {author}
  {\bibfnamefont {P.~E.}\ \bibnamefont {Barclay}},\ }\href {\doibase
  10.1103/PhysRevX.4.021052} {\bibfield  {journal} {\bibinfo  {journal} {Phys.
  Rev. X}\ }\textbf {\bibinfo {volume} {4}},\ \bibinfo {pages} {021052}
  (\bibinfo {year} {2014})}\BibitemShut {NoStop}%
\bibitem [{\citenamefont {Gao}\ and\ \citenamefont {Wang}(2021)}]{Gao:21}%
  \BibitemOpen
  \bibfield  {author} {\bibinfo {author} {\bibfnamefont {Y.-P.}\ \bibnamefont
  {Gao}}\ and\ \bibinfo {author} {\bibfnamefont {C.}~\bibnamefont {Wang}},\
  }\href {\doibase 10.1364/OE.431211} {\bibfield  {journal} {\bibinfo
  {journal} {Opt. Express}\ }\textbf {\bibinfo {volume} {29}},\ \bibinfo
  {pages} {25161} (\bibinfo {year} {2021})}\BibitemShut {NoStop}%
\bibitem [{\citenamefont {Yamamoto}\ \emph {et~al.}(2010)\citenamefont
  {Yamamoto}, \citenamefont {Friedrich}, \citenamefont {Westphal},
  \citenamefont {Go\ss{}ler}, \citenamefont {Danzmann}, \citenamefont {Somiya},
  \citenamefont {Danilishin},\ and\ \citenamefont
  {Schnabel}}]{PhysRevA.81.033849}%
  \BibitemOpen
  \bibfield  {author} {\bibinfo {author} {\bibfnamefont {K.}~\bibnamefont
  {Yamamoto}}, \bibinfo {author} {\bibfnamefont {D.}~\bibnamefont {Friedrich}},
  \bibinfo {author} {\bibfnamefont {T.}~\bibnamefont {Westphal}}, \bibinfo
  {author} {\bibfnamefont {S.}~\bibnamefont {Go\ss{}ler}}, \bibinfo {author}
  {\bibfnamefont {K.}~\bibnamefont {Danzmann}}, \bibinfo {author}
  {\bibfnamefont {K.}~\bibnamefont {Somiya}}, \bibinfo {author} {\bibfnamefont
  {S.~L.}\ \bibnamefont {Danilishin}}, \ and\ \bibinfo {author} {\bibfnamefont
  {R.}~\bibnamefont {Schnabel}},\ }\href {\doibase 10.1103/PhysRevA.81.033849}
  {\bibfield  {journal} {\bibinfo  {journal} {Phys. Rev. A}\ }\textbf {\bibinfo
  {volume} {81}},\ \bibinfo {pages} {033849} (\bibinfo {year}
  {2010})}\BibitemShut {NoStop}%
\bibitem [{\citenamefont {Tarabrin}\ \emph {et~al.}(2013)\citenamefont
  {Tarabrin}, \citenamefont {Kaufer}, \citenamefont {Khalili}, \citenamefont
  {Schnabel},\ and\ \citenamefont {Hammerer}}]{PhysRevA.88.023809}%
  \BibitemOpen
  \bibfield  {author} {\bibinfo {author} {\bibfnamefont {S.~P.}\ \bibnamefont
  {Tarabrin}}, \bibinfo {author} {\bibfnamefont {H.}~\bibnamefont {Kaufer}},
  \bibinfo {author} {\bibfnamefont {F.~Y.}\ \bibnamefont {Khalili}}, \bibinfo
  {author} {\bibfnamefont {R.}~\bibnamefont {Schnabel}}, \ and\ \bibinfo
  {author} {\bibfnamefont {K.}~\bibnamefont {Hammerer}},\ }\href {\doibase
  10.1103/PhysRevA.88.023809} {\bibfield  {journal} {\bibinfo  {journal} {Phys.
  Rev. A}\ }\textbf {\bibinfo {volume} {88}},\ \bibinfo {pages} {023809}
  (\bibinfo {year} {2013})}\BibitemShut {NoStop}%
\bibitem [{\citenamefont {Zoller}\ and\ \citenamefont
  {Gardiner}(1997)}]{zoller1997quantum}%
  \BibitemOpen
  \bibfield  {author} {\bibinfo {author} {\bibfnamefont {P.}~\bibnamefont
  {Zoller}}\ and\ \bibinfo {author} {\bibfnamefont {C.~W.}\ \bibnamefont
  {Gardiner}},\ }\href@noop {} {\enquote {\bibinfo {title} {Quantum noise in
  quantum optics: the stochastic schr\"odinger equation},}\ } (\bibinfo {year}
  {1997}),\ \Eprint {http://arxiv.org/abs/quant-ph/9702030}
  {arXiv:quant-ph/9702030 [quant-ph]} \BibitemShut {NoStop}%
\bibitem [{\citenamefont {Clerk}\ \emph {et~al.}(2010)\citenamefont {Clerk},
  \citenamefont {Devoret}, \citenamefont {Girvin}, \citenamefont {Marquardt},\
  and\ \citenamefont {Schoelkopf}}]{RevModPhys.82.1155}%
  \BibitemOpen
  \bibfield  {author} {\bibinfo {author} {\bibfnamefont {A.~A.}\ \bibnamefont
  {Clerk}}, \bibinfo {author} {\bibfnamefont {M.~H.}\ \bibnamefont {Devoret}},
  \bibinfo {author} {\bibfnamefont {S.~M.}\ \bibnamefont {Girvin}}, \bibinfo
  {author} {\bibfnamefont {F.}~\bibnamefont {Marquardt}}, \ and\ \bibinfo
  {author} {\bibfnamefont {R.~J.}\ \bibnamefont {Schoelkopf}},\ }\href
  {\doibase 10.1103/RevModPhys.82.1155} {\bibfield  {journal} {\bibinfo
  {journal} {Rev. Mod. Phys.}\ }\textbf {\bibinfo {volume} {82}},\ \bibinfo
  {pages} {1155} (\bibinfo {year} {2010})}\BibitemShut {NoStop}%
\bibitem [{\citenamefont {Gardiner}\ and\ \citenamefont
  {Collett}(1985)}]{PhysRevA.31.3761}%
  \BibitemOpen
  \bibfield  {author} {\bibinfo {author} {\bibfnamefont {C.~W.}\ \bibnamefont
  {Gardiner}}\ and\ \bibinfo {author} {\bibfnamefont {M.~J.}\ \bibnamefont
  {Collett}},\ }\href {\doibase 10.1103/PhysRevA.31.3761} {\bibfield  {journal}
  {\bibinfo  {journal} {Phys. Rev. A}\ }\textbf {\bibinfo {volume} {31}},\
  \bibinfo {pages} {3761} (\bibinfo {year} {1985})}\BibitemShut {NoStop}%
\bibitem [{\citenamefont {He}\ \emph {et~al.}(2022)\citenamefont {He},
  \citenamefont {Badshah}, \citenamefont {Song}, \citenamefont {Wang},
  \citenamefont {Liang},\ and\ \citenamefont {Su}}]{PhysRevA.105.013503}%
  \BibitemOpen
  \bibfield  {author} {\bibinfo {author} {\bibfnamefont {Q.}~\bibnamefont
  {He}}, \bibinfo {author} {\bibfnamefont {F.}~\bibnamefont {Badshah}},
  \bibinfo {author} {\bibfnamefont {Y.}~\bibnamefont {Song}}, \bibinfo {author}
  {\bibfnamefont {L.}~\bibnamefont {Wang}}, \bibinfo {author} {\bibfnamefont
  {E.}~\bibnamefont {Liang}}, \ and\ \bibinfo {author} {\bibfnamefont {S.-L.}\
  \bibnamefont {Su}},\ }\href {\doibase 10.1103/PhysRevA.105.013503} {\bibfield
   {journal} {\bibinfo  {journal} {Phys. Rev. A}\ }\textbf {\bibinfo {volume}
  {105}},\ \bibinfo {pages} {013503} (\bibinfo {year} {2022})}\BibitemShut
  {NoStop}%
\bibitem [{\citenamefont {Mehmood}\ \emph {et~al.}(2020)\citenamefont
  {Mehmood}, \citenamefont {Qamar},\ and\ \citenamefont
  {Qamar}}]{Mehmood_2020}%
  \BibitemOpen
  \bibfield  {author} {\bibinfo {author} {\bibfnamefont {A.}~\bibnamefont
  {Mehmood}}, \bibinfo {author} {\bibfnamefont {S.}~\bibnamefont {Qamar}}, \
  and\ \bibinfo {author} {\bibfnamefont {S.}~\bibnamefont {Qamar}},\ }\href
  {\doibase 10.1088/1402-4896/ab553e} {\bibfield  {journal} {\bibinfo
  {journal} {Physica Scripta}\ }\textbf {\bibinfo {volume} {95}},\ \bibinfo
  {pages} {035102} (\bibinfo {year} {2020})}\BibitemShut {NoStop}%
\bibitem [{\citenamefont {Hao}\ \emph {et~al.}(2021)\citenamefont {Hao},
  \citenamefont {Zhang}, \citenamefont {Zhou}, \citenamefont {Li},
  \citenamefont {Hou},\ and\ \citenamefont {Yi}}]{PhysRevA.103.053508}%
  \BibitemOpen
  \bibfield  {author} {\bibinfo {author} {\bibfnamefont {X.~Z.}\ \bibnamefont
  {Hao}}, \bibinfo {author} {\bibfnamefont {X.~Y.}\ \bibnamefont {Zhang}},
  \bibinfo {author} {\bibfnamefont {Y.~H.}\ \bibnamefont {Zhou}}, \bibinfo
  {author} {\bibfnamefont {W.}~\bibnamefont {Li}}, \bibinfo {author}
  {\bibfnamefont {S.~C.}\ \bibnamefont {Hou}}, \ and\ \bibinfo {author}
  {\bibfnamefont {X.~X.}\ \bibnamefont {Yi}},\ }\href {\doibase
  10.1103/PhysRevA.103.053508} {\bibfield  {journal} {\bibinfo  {journal}
  {Phys. Rev. A}\ }\textbf {\bibinfo {volume} {103}},\ \bibinfo {pages}
  {053508} (\bibinfo {year} {2021})}\BibitemShut {NoStop}%
\bibitem [{\citenamefont {Arcizet}\ \emph {et~al.}(2011)\citenamefont
  {Arcizet}, \citenamefont {Jacques}, \citenamefont {Siria}, \citenamefont
  {Poncharal}, \citenamefont {Vincent},\ and\ \citenamefont
  {Seidelin}}]{nphys.7.879}%
  \BibitemOpen
  \bibfield  {author} {\bibinfo {author} {\bibfnamefont {O.}~\bibnamefont
  {Arcizet}}, \bibinfo {author} {\bibfnamefont {V.}~\bibnamefont {Jacques}},
  \bibinfo {author} {\bibfnamefont {a.}~\bibnamefont {Siria}}, \bibinfo
  {author} {\bibfnamefont {P.}~\bibnamefont {Poncharal}}, \bibinfo {author}
  {\bibfnamefont {P.}~\bibnamefont {Vincent}}, \ and\ \bibinfo {author}
  {\bibfnamefont {S.}~\bibnamefont {Seidelin}},\ }\href {\doibase
  10.1038/nphys2070} {\bibfield  {journal} {\bibinfo  {journal} {Nature
  Physics}\ }\textbf {\bibinfo {volume} {7}},\ \bibinfo {pages} {879} (\bibinfo
  {year} {2011})}\BibitemShut {NoStop}%
\bibitem [{\citenamefont {Bamba}\ \emph {et~al.}(2011)\citenamefont {Bamba},
  \citenamefont {Imamo\ifmmode~\breve{g}\else \u{g}\fi{}lu}, \citenamefont
  {Carusotto},\ and\ \citenamefont {Ciuti}}]{PhysRevA.83.021802}%
  \BibitemOpen
  \bibfield  {author} {\bibinfo {author} {\bibfnamefont {M.}~\bibnamefont
  {Bamba}}, \bibinfo {author} {\bibfnamefont {A.}~\bibnamefont
  {Imamo\ifmmode~\breve{g}\else \u{g}\fi{}lu}}, \bibinfo {author}
  {\bibfnamefont {I.}~\bibnamefont {Carusotto}}, \ and\ \bibinfo {author}
  {\bibfnamefont {C.}~\bibnamefont {Ciuti}},\ }\href {\doibase
  10.1103/PhysRevA.83.021802} {\bibfield  {journal} {\bibinfo  {journal} {Phys.
  Rev. A}\ }\textbf {\bibinfo {volume} {83}},\ \bibinfo {pages} {021802(R)}
  (\bibinfo {year} {2011})}\BibitemShut {NoStop}%
\bibitem [{\citenamefont {Zhang}\ \emph
  {et~al.}(2015{\natexlab{a}})\citenamefont {Zhang}, \citenamefont {Cheng},
  \citenamefont {Liu},\ and\ \citenamefont {Zhou}}]{PhysRevA.91.063836}%
  \BibitemOpen
  \bibfield  {author} {\bibinfo {author} {\bibfnamefont {W.-Z.}\ \bibnamefont
  {Zhang}}, \bibinfo {author} {\bibfnamefont {J.}~\bibnamefont {Cheng}},
  \bibinfo {author} {\bibfnamefont {J.-Y.}\ \bibnamefont {Liu}}, \ and\
  \bibinfo {author} {\bibfnamefont {L.}~\bibnamefont {Zhou}},\ }\href {\doibase
  10.1103/PhysRevA.91.063836} {\bibfield  {journal} {\bibinfo  {journal} {Phys.
  Rev. A}\ }\textbf {\bibinfo {volume} {91}},\ \bibinfo {pages} {063836}
  (\bibinfo {year} {2015}{\natexlab{a}})}\BibitemShut {NoStop}%
\bibitem [{\citenamefont {Safavi-Naeini}\ and\ \citenamefont
  {Painter}(2010)}]{OE.18.014926}%
  \BibitemOpen
  \bibfield  {author} {\bibinfo {author} {\bibfnamefont {A.~H.}\ \bibnamefont
  {Safavi-Naeini}}\ and\ \bibinfo {author} {\bibfnamefont {O.}~\bibnamefont
  {Painter}},\ }\href {\doibase 10.1364/OE.18.014926} {\bibfield  {journal}
  {\bibinfo  {journal} {Optics Express}\ }\textbf {\bibinfo {volume} {18}},\
  \bibinfo {pages} {14926} (\bibinfo {year} {2010})}\BibitemShut {NoStop}%
\bibitem [{\citenamefont {Cai}\ \emph {et~al.}(2020)\citenamefont {Cai},
  \citenamefont {Pan},\ and\ \citenamefont {Hu}}]{OLE.127.105968}%
  \BibitemOpen
  \bibfield  {author} {\bibinfo {author} {\bibfnamefont {L.}~\bibnamefont
  {Cai}}, \bibinfo {author} {\bibfnamefont {J.}~\bibnamefont {Pan}}, \ and\
  \bibinfo {author} {\bibfnamefont {S.}~\bibnamefont {Hu}},\ }\href {\doibase
  10.1016/j.optlaseng.2019.105968} {\bibfield  {journal} {\bibinfo  {journal}
  {Optics and Lasers in Engineering}\ }\textbf {\bibinfo {volume} {127}},\
  \bibinfo {pages} {105968} (\bibinfo {year} {2020})}\BibitemShut {NoStop}%
\bibitem [{\citenamefont {Yuan}\ \emph {et~al.}(2015)\citenamefont {Yuan},
  \citenamefont {Singh}, \citenamefont {Blanter},\ and\ \citenamefont
  {Steele}}]{Ncomms.6.8491}%
  \BibitemOpen
  \bibfield  {author} {\bibinfo {author} {\bibfnamefont {M.}~\bibnamefont
  {Yuan}}, \bibinfo {author} {\bibfnamefont {V.}~\bibnamefont {Singh}},
  \bibinfo {author} {\bibfnamefont {Y.~M.}\ \bibnamefont {Blanter}}, \ and\
  \bibinfo {author} {\bibfnamefont {G.~A.}\ \bibnamefont {Steele}},\ }\href
  {\doibase 10.1038/ncomms9491} {\bibfield  {journal} {\bibinfo  {journal}
  {Nature Communications}\ }\textbf {\bibinfo {volume} {6}},\ \bibinfo {pages}
  {8491} (\bibinfo {year} {2015})}\BibitemShut {NoStop}%
\bibitem [{\citenamefont {Arcizet}\ \emph {et~al.}(2006)\citenamefont
  {Arcizet}, \citenamefont {Cohadon}, \citenamefont {Briant}, \citenamefont
  {Pinard}, \citenamefont {Heidmann}, \citenamefont {Mackowski}, \citenamefont
  {Michel}, \citenamefont {Pinard}, \citenamefont
  {Fran\ifmmode~\mbox{\c{c}}\else \c{c}\fi{}ais},\ and\ \citenamefont
  {Rousseau}}]{PhysRevLett.97.133601}%
  \BibitemOpen
  \bibfield  {author} {\bibinfo {author} {\bibfnamefont {O.}~\bibnamefont
  {Arcizet}}, \bibinfo {author} {\bibfnamefont {P.-F.}\ \bibnamefont
  {Cohadon}}, \bibinfo {author} {\bibfnamefont {T.}~\bibnamefont {Briant}},
  \bibinfo {author} {\bibfnamefont {M.}~\bibnamefont {Pinard}}, \bibinfo
  {author} {\bibfnamefont {A.}~\bibnamefont {Heidmann}}, \bibinfo {author}
  {\bibfnamefont {J.-M.}\ \bibnamefont {Mackowski}}, \bibinfo {author}
  {\bibfnamefont {C.}~\bibnamefont {Michel}}, \bibinfo {author} {\bibfnamefont
  {L.}~\bibnamefont {Pinard}}, \bibinfo {author} {\bibfnamefont
  {O.}~\bibnamefont {Fran\ifmmode~\mbox{\c{c}}\else \c{c}\fi{}ais}}, \ and\
  \bibinfo {author} {\bibfnamefont {L.}~\bibnamefont {Rousseau}},\ }\href
  {\doibase 10.1103/PhysRevLett.97.133601} {\bibfield  {journal} {\bibinfo
  {journal} {Phys. Rev. Lett.}\ }\textbf {\bibinfo {volume} {97}},\ \bibinfo
  {pages} {133601} (\bibinfo {year} {2006})}\BibitemShut {NoStop}%
\bibitem [{\citenamefont {Teufel}\ \emph
  {et~al.}(2011{\natexlab{b}})\citenamefont {Teufel}, \citenamefont {Donner},
  \citenamefont {Li}, \citenamefont {Harlow}, \citenamefont {Allman},
  \citenamefont {Cicak}, \citenamefont {Sirois}, \citenamefont {Whittaker},
  \citenamefont {Lehnert},\ and\ \citenamefont {Simmonds}}]{nature.475.359}%
  \BibitemOpen
  \bibfield  {author} {\bibinfo {author} {\bibfnamefont {J.~D.}\ \bibnamefont
  {Teufel}}, \bibinfo {author} {\bibfnamefont {T.}~\bibnamefont {Donner}},
  \bibinfo {author} {\bibfnamefont {D.}~\bibnamefont {Li}}, \bibinfo {author}
  {\bibfnamefont {J.~W.}\ \bibnamefont {Harlow}}, \bibinfo {author}
  {\bibfnamefont {M.~S.}\ \bibnamefont {Allman}}, \bibinfo {author}
  {\bibfnamefont {K.}~\bibnamefont {Cicak}}, \bibinfo {author} {\bibfnamefont
  {A.~J.}\ \bibnamefont {Sirois}}, \bibinfo {author} {\bibfnamefont {J.~D.}\
  \bibnamefont {Whittaker}}, \bibinfo {author} {\bibfnamefont {K.~W.}\
  \bibnamefont {Lehnert}}, \ and\ \bibinfo {author} {\bibfnamefont {R.~W.}\
  \bibnamefont {Simmonds}},\ }\href {\doibase 10.1038/nature10261} {\bibfield
  {journal} {\bibinfo  {journal} {Nature}\ }\textbf {\bibinfo {volume} {475}},\
  \bibinfo {pages} {359} (\bibinfo {year} {2011}{\natexlab{b}})}\BibitemShut
  {NoStop}%
\bibitem [{\citenamefont {Park}\ and\ \citenamefont
  {Wang}(2009)}]{nphys.5.489}%
  \BibitemOpen
  \bibfield  {author} {\bibinfo {author} {\bibfnamefont {Y.-S.}\ \bibnamefont
  {Park}}\ and\ \bibinfo {author} {\bibfnamefont {H.}~\bibnamefont {Wang}},\
  }\href {\doibase 10.1038/nphys1303} {\bibfield  {journal} {\bibinfo
  {journal} {Nature Physics}\ }\textbf {\bibinfo {volume} {5}},\ \bibinfo
  {pages} {489} (\bibinfo {year} {2009})}\BibitemShut {NoStop}%
\bibitem [{\citenamefont {Chen}\ \emph {et~al.}(2021)\citenamefont {Chen},
  \citenamefont {Zhang}, \citenamefont {Shen}, \citenamefont {Zou},
  \citenamefont {Guo},\ and\ \citenamefont {Dong}}]{PhysRevLett.126.123603}%
  \BibitemOpen
  \bibfield  {author} {\bibinfo {author} {\bibfnamefont {Y.}~\bibnamefont
  {Chen}}, \bibinfo {author} {\bibfnamefont {Y.-L.}\ \bibnamefont {Zhang}},
  \bibinfo {author} {\bibfnamefont {Z.}~\bibnamefont {Shen}}, \bibinfo {author}
  {\bibfnamefont {C.-L.}\ \bibnamefont {Zou}}, \bibinfo {author} {\bibfnamefont
  {G.-C.}\ \bibnamefont {Guo}}, \ and\ \bibinfo {author} {\bibfnamefont
  {C.-H.}\ \bibnamefont {Dong}},\ }\href {\doibase
  10.1103/PhysRevLett.126.123603} {\bibfield  {journal} {\bibinfo  {journal}
  {Phys. Rev. Lett.}\ }\textbf {\bibinfo {volume} {126}},\ \bibinfo {pages}
  {123603} (\bibinfo {year} {2021})}\BibitemShut {NoStop}%
\bibitem [{\citenamefont {Teufel}\ \emph
  {et~al.}(2011{\natexlab{c}})\citenamefont {Teufel}, \citenamefont {Li},
  \citenamefont {Allman}, \citenamefont {Cicak}, \citenamefont {Sirois},
  \citenamefont {Whittaker},\ and\ \citenamefont {Simmonds}}]{nature.471.204}%
  \BibitemOpen
  \bibfield  {author} {\bibinfo {author} {\bibfnamefont {J.~D.}\ \bibnamefont
  {Teufel}}, \bibinfo {author} {\bibfnamefont {D.}~\bibnamefont {Li}}, \bibinfo
  {author} {\bibfnamefont {M.~S.}\ \bibnamefont {Allman}}, \bibinfo {author}
  {\bibfnamefont {K.}~\bibnamefont {Cicak}}, \bibinfo {author} {\bibfnamefont
  {A.~J.}\ \bibnamefont {Sirois}}, \bibinfo {author} {\bibfnamefont {J.~D.}\
  \bibnamefont {Whittaker}}, \ and\ \bibinfo {author} {\bibfnamefont {R.~W.}\
  \bibnamefont {Simmonds}},\ }\href {\doibase 10.1038/nature09898} {\bibfield
  {journal} {\bibinfo  {journal} {Nature}\ }\textbf {\bibinfo {volume} {471}},\
  \bibinfo {pages} {204} (\bibinfo {year} {2011}{\natexlab{c}})}\BibitemShut
  {NoStop}%
\bibitem [{\citenamefont {Brooks}\ \emph {et~al.}(2012)\citenamefont {Brooks},
  \citenamefont {Botter}, \citenamefont {Schreppler}, \citenamefont {Purdy},
  \citenamefont {Brahms},\ and\ \citenamefont {Stamper-Kurn}}]{Nature.488.476}%
  \BibitemOpen
  \bibfield  {author} {\bibinfo {author} {\bibfnamefont {D.~W.~C.}\
  \bibnamefont {Brooks}}, \bibinfo {author} {\bibfnamefont {T.}~\bibnamefont
  {Botter}}, \bibinfo {author} {\bibfnamefont {S.}~\bibnamefont {Schreppler}},
  \bibinfo {author} {\bibfnamefont {T.~P.}\ \bibnamefont {Purdy}}, \bibinfo
  {author} {\bibfnamefont {N.}~\bibnamefont {Brahms}}, \ and\ \bibinfo {author}
  {\bibfnamefont {D.~M.}\ \bibnamefont {Stamper-Kurn}},\ }\href {\doibase
  10.1038/nature11325} {\bibfield  {journal} {\bibinfo  {journal} {Nature}\
  }\textbf {\bibinfo {volume} {488}},\ \bibinfo {pages} {476} (\bibinfo {year}
  {2012})}\BibitemShut {NoStop}%
\bibitem [{\citenamefont {Liao}\ and\ \citenamefont
  {Nori}(2013)}]{PhysRevA.88.023853}%
  \BibitemOpen
  \bibfield  {author} {\bibinfo {author} {\bibfnamefont {J.-Q.}\ \bibnamefont
  {Liao}}\ and\ \bibinfo {author} {\bibfnamefont {F.}~\bibnamefont {Nori}},\
  }\href {\doibase 10.1103/PhysRevA.88.023853} {\bibfield  {journal} {\bibinfo
  {journal} {Phys. Rev. A}\ }\textbf {\bibinfo {volume} {88}},\ \bibinfo
  {pages} {023853} (\bibinfo {year} {2013})}\BibitemShut {NoStop}%
\bibitem [{\citenamefont {Hou}\ \emph {et~al.}(2019)\citenamefont {Hou},
  \citenamefont {Zhu}, \citenamefont {Yang},\ and\ \citenamefont
  {Agarwal}}]{PhysRevA.100.063817}%
  \BibitemOpen
  \bibfield  {author} {\bibinfo {author} {\bibfnamefont {K.}~\bibnamefont
  {Hou}}, \bibinfo {author} {\bibfnamefont {C.~J.}\ \bibnamefont {Zhu}},
  \bibinfo {author} {\bibfnamefont {Y.~P.}\ \bibnamefont {Yang}}, \ and\
  \bibinfo {author} {\bibfnamefont {G.~S.}\ \bibnamefont {Agarwal}},\ }\href
  {\doibase 10.1103/PhysRevA.100.063817} {\bibfield  {journal} {\bibinfo
  {journal} {Phys. Rev. A}\ }\textbf {\bibinfo {volume} {100}},\ \bibinfo
  {pages} {063817} (\bibinfo {year} {2019})}\BibitemShut {NoStop}%
\bibitem [{\citenamefont {Flayac}\ and\ \citenamefont
  {Savona}(2017)}]{PhysRevA.96.053810}%
  \BibitemOpen
  \bibfield  {author} {\bibinfo {author} {\bibfnamefont {H.}~\bibnamefont
  {Flayac}}\ and\ \bibinfo {author} {\bibfnamefont {V.}~\bibnamefont
  {Savona}},\ }\href {\doibase 10.1103/PhysRevA.96.053810} {\bibfield
  {journal} {\bibinfo  {journal} {Phys. Rev. A}\ }\textbf {\bibinfo {volume}
  {96}},\ \bibinfo {pages} {053810} (\bibinfo {year} {2017})}\BibitemShut
  {NoStop}%
\bibitem [{\citenamefont {Lemonde}\ \emph {et~al.}(2014)\citenamefont
  {Lemonde}, \citenamefont {Didier},\ and\ \citenamefont
  {Clerk}}]{PhysRevA.90.063824}%
  \BibitemOpen
  \bibfield  {author} {\bibinfo {author} {\bibfnamefont {M.-A.}\ \bibnamefont
  {Lemonde}}, \bibinfo {author} {\bibfnamefont {N.}~\bibnamefont {Didier}}, \
  and\ \bibinfo {author} {\bibfnamefont {A.~A.}\ \bibnamefont {Clerk}},\ }\href
  {\doibase 10.1103/PhysRevA.90.063824} {\bibfield  {journal} {\bibinfo
  {journal} {Phys. Rev. A}\ }\textbf {\bibinfo {volume} {90}},\ \bibinfo
  {pages} {063824} (\bibinfo {year} {2014})}\BibitemShut {NoStop}%
\bibitem [{\citenamefont {Sarma}\ and\ \citenamefont
  {Sarma}(2018)}]{PhysRevA.98.013826}%
  \BibitemOpen
  \bibfield  {author} {\bibinfo {author} {\bibfnamefont {B.}~\bibnamefont
  {Sarma}}\ and\ \bibinfo {author} {\bibfnamefont {A.~K.}\ \bibnamefont
  {Sarma}},\ }\href {\doibase 10.1103/PhysRevA.98.013826} {\bibfield  {journal}
  {\bibinfo  {journal} {Phys. Rev. A}\ }\textbf {\bibinfo {volume} {98}},\
  \bibinfo {pages} {013826} (\bibinfo {year} {2018})}\BibitemShut {NoStop}%
\bibitem [{\citenamefont {Shen}\ \emph {et~al.}(2018)\citenamefont {Shen},
  \citenamefont {Shang}, \citenamefont {Zhou},\ and\ \citenamefont
  {Yi}}]{PhysRevA.98.023856}%
  \BibitemOpen
  \bibfield  {author} {\bibinfo {author} {\bibfnamefont {H.~Z.}\ \bibnamefont
  {Shen}}, \bibinfo {author} {\bibfnamefont {C.}~\bibnamefont {Shang}},
  \bibinfo {author} {\bibfnamefont {Y.~H.}\ \bibnamefont {Zhou}}, \ and\
  \bibinfo {author} {\bibfnamefont {X.~X.}\ \bibnamefont {Yi}},\ }\href
  {\doibase 10.1103/PhysRevA.98.023856} {\bibfield  {journal} {\bibinfo
  {journal} {Phys. Rev. A}\ }\textbf {\bibinfo {volume} {98}},\ \bibinfo
  {pages} {023856} (\bibinfo {year} {2018})}\BibitemShut {NoStop}%
\bibitem [{\citenamefont {Wang}\ \emph {et~al.}(2013)\citenamefont {Wang},
  \citenamefont {Zhou}, \citenamefont {Guo}, \citenamefont {Zhang},
  \citenamefont {Evers},\ and\ \citenamefont {Zhu}}]{PhysRevLett.110.093901}%
  \BibitemOpen
  \bibfield  {author} {\bibinfo {author} {\bibfnamefont {D.-W.}\ \bibnamefont
  {Wang}}, \bibinfo {author} {\bibfnamefont {H.-T.}\ \bibnamefont {Zhou}},
  \bibinfo {author} {\bibfnamefont {M.-J.}\ \bibnamefont {Guo}}, \bibinfo
  {author} {\bibfnamefont {J.-X.}\ \bibnamefont {Zhang}}, \bibinfo {author}
  {\bibfnamefont {J.}~\bibnamefont {Evers}}, \ and\ \bibinfo {author}
  {\bibfnamefont {S.-Y.}\ \bibnamefont {Zhu}},\ }\href {\doibase
  10.1103/PhysRevLett.110.093901} {\bibfield  {journal} {\bibinfo  {journal}
  {Phys. Rev. Lett.}\ }\textbf {\bibinfo {volume} {110}},\ \bibinfo {pages}
  {093901} (\bibinfo {year} {2013})}\BibitemShut {NoStop}%
\bibitem [{\citenamefont {Blakesley}\ \emph {et~al.}(2005)\citenamefont
  {Blakesley}, \citenamefont {See}, \citenamefont {Shields}, \citenamefont
  {Kardyna\l{}}, \citenamefont {Atkinson}, \citenamefont {Farrer},\ and\
  \citenamefont {Ritchie}}]{PhysRevLett.94.067401}%
  \BibitemOpen
  \bibfield  {author} {\bibinfo {author} {\bibfnamefont {J.~C.}\ \bibnamefont
  {Blakesley}}, \bibinfo {author} {\bibfnamefont {P.}~\bibnamefont {See}},
  \bibinfo {author} {\bibfnamefont {A.~J.}\ \bibnamefont {Shields}}, \bibinfo
  {author} {\bibfnamefont {B.~E.}\ \bibnamefont {Kardyna\l{}}}, \bibinfo
  {author} {\bibfnamefont {P.}~\bibnamefont {Atkinson}}, \bibinfo {author}
  {\bibfnamefont {I.}~\bibnamefont {Farrer}}, \ and\ \bibinfo {author}
  {\bibfnamefont {D.~A.}\ \bibnamefont {Ritchie}},\ }\href {\doibase
  10.1103/PhysRevLett.94.067401} {\bibfield  {journal} {\bibinfo  {journal}
  {Phys. Rev. Lett.}\ }\textbf {\bibinfo {volume} {94}},\ \bibinfo {pages}
  {067401} (\bibinfo {year} {2005})}\BibitemShut {NoStop}%
\bibitem [{\citenamefont {Lin}\ and\ \citenamefont
  {Li}(2009)}]{PhysRevA.79.022301}%
  \BibitemOpen
  \bibfield  {author} {\bibinfo {author} {\bibfnamefont {Q.}~\bibnamefont
  {Lin}}\ and\ \bibinfo {author} {\bibfnamefont {J.}~\bibnamefont {Li}},\
  }\href {\doibase 10.1103/PhysRevA.79.022301} {\bibfield  {journal} {\bibinfo
  {journal} {Phys. Rev. A}\ }\textbf {\bibinfo {volume} {79}},\ \bibinfo
  {pages} {022301} (\bibinfo {year} {2009})}\BibitemShut {NoStop}%
\bibitem [{\citenamefont {Zhang}\ \emph
  {et~al.}(2015{\natexlab{b}})\citenamefont {Zhang}, \citenamefont {Cheng},\
  and\ \citenamefont {Zhou}}]{JPB.48.015502}%
  \BibitemOpen
  \bibfield  {author} {\bibinfo {author} {\bibfnamefont {W.-Z.}\ \bibnamefont
  {Zhang}}, \bibinfo {author} {\bibfnamefont {J.}~\bibnamefont {Cheng}}, \ and\
  \bibinfo {author} {\bibfnamefont {L.}~\bibnamefont {Zhou}},\ }\href {\doibase
  10.1088/0953-4075/48/1/015502} {\bibfield  {journal} {\bibinfo  {journal}
  {Journal of Physics B: Atomic, Molecular and Optical Physics}\ }\textbf
  {\bibinfo {volume} {48}},\ \bibinfo {pages} {015502} (\bibinfo {year}
  {2015}{\natexlab{b}})}\BibitemShut {NoStop}%
\bibitem [{\citenamefont {Asjad}\ \emph {et~al.}(2015)\citenamefont {Asjad},
  \citenamefont {Tombesi},\ and\ \citenamefont {Vitali}}]{OE.23.7786}%
  \BibitemOpen
  \bibfield  {author} {\bibinfo {author} {\bibfnamefont {M.}~\bibnamefont
  {Asjad}}, \bibinfo {author} {\bibfnamefont {P.}~\bibnamefont {Tombesi}}, \
  and\ \bibinfo {author} {\bibfnamefont {D.}~\bibnamefont {Vitali}},\ }\href
  {\doibase 10.1364/OE.23.007786} {\bibfield  {journal} {\bibinfo  {journal}
  {Optics Express}\ }\textbf {\bibinfo {volume} {23}},\ \bibinfo {pages} {7786}
  (\bibinfo {year} {2015})}\BibitemShut {NoStop}%
\bibitem [{\citenamefont {Mari}\ \emph {et~al.}(2013)\citenamefont {Mari},
  \citenamefont {Farace}, \citenamefont {Didier}, \citenamefont {Giovannetti},\
  and\ \citenamefont {Fazio}}]{PhysRevLett.111.103605}%
  \BibitemOpen
  \bibfield  {author} {\bibinfo {author} {\bibfnamefont {A.}~\bibnamefont
  {Mari}}, \bibinfo {author} {\bibfnamefont {A.}~\bibnamefont {Farace}},
  \bibinfo {author} {\bibfnamefont {N.}~\bibnamefont {Didier}}, \bibinfo
  {author} {\bibfnamefont {V.}~\bibnamefont {Giovannetti}}, \ and\ \bibinfo
  {author} {\bibfnamefont {R.}~\bibnamefont {Fazio}},\ }\href {\doibase
  10.1103/PhysRevLett.111.103605} {\bibfield  {journal} {\bibinfo  {journal}
  {Phys. Rev. Lett.}\ }\textbf {\bibinfo {volume} {111}},\ \bibinfo {pages}
  {103605} (\bibinfo {year} {2013})}\BibitemShut {NoStop}%
\bibitem [{\citenamefont {Li}\ \emph {et~al.}(2017)\citenamefont {Li},
  \citenamefont {Li},\ and\ \citenamefont {Song}}]{PhysRevE.95.022204}%
  \BibitemOpen
  \bibfield  {author} {\bibinfo {author} {\bibfnamefont {W.}~\bibnamefont
  {Li}}, \bibinfo {author} {\bibfnamefont {C.}~\bibnamefont {Li}}, \ and\
  \bibinfo {author} {\bibfnamefont {H.}~\bibnamefont {Song}},\ }\href {\doibase
  10.1103/PhysRevE.95.022204} {\bibfield  {journal} {\bibinfo  {journal} {Phys.
  Rev. E}\ }\textbf {\bibinfo {volume} {95}},\ \bibinfo {pages} {022204}
  (\bibinfo {year} {2017})}\BibitemShut {NoStop}%
\bibitem [{\citenamefont {Walter}\ \emph {et~al.}(2014)\citenamefont {Walter},
  \citenamefont {Nunnenkamp},\ and\ \citenamefont
  {Bruder}}]{PhysRevLett.112.094102}%
  \BibitemOpen
  \bibfield  {author} {\bibinfo {author} {\bibfnamefont {S.}~\bibnamefont
  {Walter}}, \bibinfo {author} {\bibfnamefont {A.}~\bibnamefont {Nunnenkamp}},
  \ and\ \bibinfo {author} {\bibfnamefont {C.}~\bibnamefont {Bruder}},\ }\href
  {\doibase 10.1103/PhysRevLett.112.094102} {\bibfield  {journal} {\bibinfo
  {journal} {Phys. Rev. Lett.}\ }\textbf {\bibinfo {volume} {112}},\ \bibinfo
  {pages} {094102} (\bibinfo {year} {2014})}\BibitemShut {NoStop}%
\bibitem [{\citenamefont {Zhang}\ \emph {et~al.}(2012)\citenamefont {Zhang},
  \citenamefont {Wiederhecker}, \citenamefont {Manipatruni}, \citenamefont
  {Barnard}, \citenamefont {McEuen},\ and\ \citenamefont
  {Lipson}}]{PhysRevLett.109.233906}%
  \BibitemOpen
  \bibfield  {author} {\bibinfo {author} {\bibfnamefont {M.}~\bibnamefont
  {Zhang}}, \bibinfo {author} {\bibfnamefont {G.~S.}\ \bibnamefont
  {Wiederhecker}}, \bibinfo {author} {\bibfnamefont {S.}~\bibnamefont
  {Manipatruni}}, \bibinfo {author} {\bibfnamefont {A.}~\bibnamefont
  {Barnard}}, \bibinfo {author} {\bibfnamefont {P.}~\bibnamefont {McEuen}}, \
  and\ \bibinfo {author} {\bibfnamefont {M.}~\bibnamefont {Lipson}},\ }\href
  {\doibase 10.1103/PhysRevLett.109.233906} {\bibfield  {journal} {\bibinfo
  {journal} {Phys. Rev. Lett.}\ }\textbf {\bibinfo {volume} {109}},\ \bibinfo
  {pages} {233906} (\bibinfo {year} {2012})}\BibitemShut {NoStop}%
\bibitem [{\citenamefont {L\"u}\ \emph {et~al.}(2015)\citenamefont {L\"u},
  \citenamefont {Jing}, \citenamefont {Ma},\ and\ \citenamefont
  {Wu}}]{PhysRevLett.114.253601}%
  \BibitemOpen
  \bibfield  {author} {\bibinfo {author} {\bibfnamefont {X.-Y.}\ \bibnamefont
  {L\"u}}, \bibinfo {author} {\bibfnamefont {H.}~\bibnamefont {Jing}}, \bibinfo
  {author} {\bibfnamefont {J.-Y.}\ \bibnamefont {Ma}}, \ and\ \bibinfo {author}
  {\bibfnamefont {Y.}~\bibnamefont {Wu}},\ }\href {\doibase
  10.1103/PhysRevLett.114.253601} {\bibfield  {journal} {\bibinfo  {journal}
  {Phys. Rev. Lett.}\ }\textbf {\bibinfo {volume} {114}},\ \bibinfo {pages}
  {253601} (\bibinfo {year} {2015})}\BibitemShut {NoStop}%
\bibitem [{\citenamefont {Rozenbaum}\ \emph {et~al.}(2020)\citenamefont
  {Rozenbaum}, \citenamefont {Bunimovich},\ and\ \citenamefont
  {Galitski}}]{PhysRevLett.125.014101}%
  \BibitemOpen
  \bibfield  {author} {\bibinfo {author} {\bibfnamefont {E.~B.}\ \bibnamefont
  {Rozenbaum}}, \bibinfo {author} {\bibfnamefont {L.~A.}\ \bibnamefont
  {Bunimovich}}, \ and\ \bibinfo {author} {\bibfnamefont {V.}~\bibnamefont
  {Galitski}},\ }\href {\doibase 10.1103/PhysRevLett.125.014101} {\bibfield
  {journal} {\bibinfo  {journal} {Phys. Rev. Lett.}\ }\textbf {\bibinfo
  {volume} {125}},\ \bibinfo {pages} {014101} (\bibinfo {year}
  {2020})}\BibitemShut {NoStop}%
\bibitem [{\citenamefont {Chaudhury}\ \emph {et~al.}(2009)\citenamefont
  {Chaudhury}, \citenamefont {Smith}, \citenamefont {Anderson}, \citenamefont
  {Ghose},\ and\ \citenamefont {Jessen}}]{nature.461.768}%
  \BibitemOpen
  \bibfield  {author} {\bibinfo {author} {\bibfnamefont {S.}~\bibnamefont
  {Chaudhury}}, \bibinfo {author} {\bibfnamefont {A.}~\bibnamefont {Smith}},
  \bibinfo {author} {\bibfnamefont {B.~E.}\ \bibnamefont {Anderson}}, \bibinfo
  {author} {\bibfnamefont {S.}~\bibnamefont {Ghose}}, \ and\ \bibinfo {author}
  {\bibfnamefont {P.~S.}\ \bibnamefont {Jessen}},\ }\href {\doibase
  10.1038/nature08396} {\bibfield  {journal} {\bibinfo  {journal} {Nature}\
  }\textbf {\bibinfo {volume} {461}},\ \bibinfo {pages} {768} (\bibinfo {year}
  {2009})}\BibitemShut {NoStop}%
\end{thebibliography}%
\end{document}